\documentclass[reprint,superscriptaddress,amsmath,amssymb,prl,showpacs,floatfix]{revtex4-1}
\usepackage{cases}
\usepackage{amsmath}
\usepackage{amssymb}
\usepackage{amsfonts}
\usepackage{amssymb}
\usepackage{dcolumn}
\usepackage{bm}
\usepackage{hyperref}
\usepackage{graphicx}
\usepackage{upgreek}
\usepackage{subfigure}
\usepackage{color}

\begin{document}

\title{Kilohertz electron paramagnetic resonance spectroscopy of single nitrogen centers at zero magnetic field}


\author{Fei Kong}
\altaffiliation{These authors contributed equally to this work.}
\affiliation{Hefei National Laboratory for Physical Sciences at the Microscale and Department of Modern Physics, University of Science and Technology of China (USTC), Hefei 230026, China}
\affiliation{CAS Key Laboratory of Microscale Magnetic Resonance, USTC, Hefei 230026, China}
\affiliation{Synergetic Innovation Center of Quantum Information and Quantum Physics, USTC, Hefei 230026, China}

\author{Pengju Zhao}
\altaffiliation{These authors contributed equally to this work.}
\affiliation{Hefei National Laboratory for Physical Sciences at the Microscale and Department of Modern Physics, University of Science and Technology of China (USTC), Hefei 230026, China}
\affiliation{CAS Key Laboratory of Microscale Magnetic Resonance, USTC, Hefei 230026, China}
\affiliation{Synergetic Innovation Center of Quantum Information and Quantum Physics, USTC, Hefei 230026, China}

\author{Pei Yu}
\altaffiliation{These authors contributed equally to this work.}
\affiliation{Hefei National Laboratory for Physical Sciences at the Microscale and Department of Modern Physics, University of Science and Technology of China (USTC), Hefei 230026, China}
\affiliation{CAS Key Laboratory of Microscale Magnetic Resonance, USTC, Hefei 230026, China}
\affiliation{Synergetic Innovation Center of Quantum Information and Quantum Physics, USTC, Hefei 230026, China}

\author{Zhuoyang Qin}
\affiliation{Hefei National Laboratory for Physical Sciences at the Microscale and Department of Modern Physics, University of Science and Technology of China (USTC), Hefei 230026, China}
\affiliation{CAS Key Laboratory of Microscale Magnetic Resonance, USTC, Hefei 230026, China}
\affiliation{Synergetic Innovation Center of Quantum Information and Quantum Physics, USTC, Hefei 230026, China}

\author{Zhehua Huang}
\affiliation{Hefei National Laboratory for Physical Sciences at the Microscale and Department of Modern Physics, University of Science and Technology of China (USTC), Hefei 230026, China}
\affiliation{CAS Key Laboratory of Microscale Magnetic Resonance, USTC, Hefei 230026, China}
\affiliation{Synergetic Innovation Center of Quantum Information and Quantum Physics, USTC, Hefei 230026, China}

\author{Zhecheng Wang}
\affiliation{Hefei National Laboratory for Physical Sciences at the Microscale and Department of Modern Physics, University of Science and Technology of China (USTC), Hefei 230026, China}
\affiliation{CAS Key Laboratory of Microscale Magnetic Resonance, USTC, Hefei 230026, China}
\affiliation{Synergetic Innovation Center of Quantum Information and Quantum Physics, USTC, Hefei 230026, China}

\author{Mengqi Wang}
\affiliation{Hefei National Laboratory for Physical Sciences at the Microscale and Department of Modern Physics, University of Science and Technology of China (USTC), Hefei 230026, China}
\affiliation{CAS Key Laboratory of Microscale Magnetic Resonance, USTC, Hefei 230026, China}
\affiliation{Synergetic Innovation Center of Quantum Information and Quantum Physics, USTC, Hefei 230026, China}

\author{Fazhan Shi}
\email{fzshi@ustc.edu.cn}
\affiliation{Hefei National Laboratory for Physical Sciences at the Microscale and Department of Modern Physics, University of Science and Technology of China (USTC), Hefei 230026, China}
\affiliation{CAS Key Laboratory of Microscale Magnetic Resonance, USTC, Hefei 230026, China}
\affiliation{Synergetic Innovation Center of Quantum Information and Quantum Physics, USTC, Hefei 230026, China}

\author{Jiangfeng Du}
\email{djf@ustc.edu.cn}
\affiliation{Hefei National Laboratory for Physical Sciences at the Microscale and Department of Modern Physics, University of Science and Technology of China (USTC), Hefei 230026, China}
\affiliation{CAS Key Laboratory of Microscale Magnetic Resonance, USTC, Hefei 230026, China}
\affiliation{Synergetic Innovation Center of Quantum Information and Quantum Physics, USTC, Hefei 230026, China}

\begin{abstract}
Electron paramagnetic resonance spectroscopy (EPR) is among the most important analytical tools in physics, chemistry, and biology. The emergence of nitrogen-vacancy (NV) centers in diamond, serving as an atomic-sized magnetometer, has promoted this technique to single-spin level, even under ambient conditions. Despite the enormous progress in spatial resolution, the current megahertz spectral resolution is still insufficient to resolve key heterogeneous molecular information. A major challenge is the short coherence times of the sample electron spins. Here, we address this challenge by employing a magnetic noise-insensitive transition between states of different symmetry. We demonstrate a 27-fold narrower spectrum of single substitutional nitrogen (P1) centers in diamond with linewidth of several kilohertz, and then some weak couplings can be resolved. Those results show both spatial and spectral advances of NV center-based EPR, and provide a route towards analytical (EPR) spectroscopy at single-molecule level.
\end{abstract}

\maketitle


Electron paramagnetic resonance (EPR) spectroscopy, a technique for studying paramagnetic targets, is an indispensable component of magnetic resonance spectroscopy for investigations of molecular structures and fast dynamics \cite{Borbat2001}. An important goal of this technique is to extract precise information from small volume samples \cite{Artzi2015}, which requires both high spatial and high spectral resolution. In the past decades, intensive efforts have been devoted to promote the spatial resolution, and ultimately single-spin EPR has been realized by various approaches, such as magnetic resonance force microscopy \cite{Rugar2004}, scanning tunneling microscopy \cite{Baumann2015}, and nitrogen-vacancy (NV) center-based EPR spectroscopy \cite{Grinolds2013,Shi2015,Shi2018}. Among them, the NV center is more promising for biological applications because of the compatibility with ambient conditions \cite{Grinolds2013,Shi2015,Shi2018}. However, the current spectral resolution of those techniques is on the orders of megahertz \cite{Rugar2004,Baumann2015,Grinolds2013,Shi2015,Shi2018,Schlipf2017,Grotz2011,Mamin2012,Hall2016,Kong2018}. It is insufficient to resolve molecules of slightly different structures \cite{Martorana2014} or local polarity profiles \cite{Kurad2003}. For example, the distance between paramagnetic centers studied by EPR techniques is in the range of 1.8 to 6 nm \cite{Jeschke2012}, corresponding to the dipolar coupling strength of megahertz to submegahertz, the variations of which induced by conformation changes will be even smaller.

The line broadening is generally attributed to the limited spin-state lifetime of sensors and the decoherence of target spins. The former one has recently been overcome by using quantum memories \cite{Laraoui2013,Zaiser2016}. The latter, which is more fundamental \cite{Sarkar2010}, arises from the magnetic couplings to bath spins. To address spin decoherence, a simple but powerful strategy is the use of particular spin states that are naturally insensitive to external perturbations. This phenomenon exists in various physical systems, for example, so-called `clock transitions' in trapped ions \cite{Bollinger1985,Harty2014} and phosphorus donors in silicon \cite{Wolfowicz2013,Morse2018}, or transitions between long-lived states in nuclear spin resonance spectroscopy \cite{Sarkar2010,Emondts2014}. The line narrowing phenomenon has also been found long ago in conventional zero-field EPR spectroscopy \cite{Cole1963,Erickson1966,Steven1984}. Nonetheless, the detection via inductive pickup suffers from the very small thermal polarization at zero field, and thus requires a large amount of samples \cite{Bramley1983}, which prevents practical applications of zero-field EPR. Fortunately, the NV center is a good magnetic sensor at zero magnetic field \cite{Zheng2019}. In a previous work \cite{Kong2018}, we have found that the NV center is a promising sensor for zero-field EPR without loss of sensitivity, because the statistical fluctuations of the spin polarization, which do not depend on the magnetic field, dominate in its nanoscale detection volumes rather than thermal polarization.

Here, we show the magnetic noise-insensitive transition can be observed by a single NV center. To investigate the spectral figure, we develop a correlation method for zero-field EPR spectroscopy. We demonstrate the high-resolution nature on single substitutional nitrogen centers (P1 centers) in diamond, which are electron spins with coupling to $^{15}$N nuclear spins. The linewidth of the observed spectra is as narrow as 8.6 kHz, which is 27-fold improvement comparing with the ordinary spectrum. Our results show not only the spatial but also the spectral advances of NV centers in EPR detections.

Our model consists of an optically probed NV center and a dark electron spin (Fig.~1A), which serve as the magnetic sensor and the target, respectively. In general, the target spin can be any spin-half electron spins ($S=1/2$) with hyperfine coupling to a nearby spin half-integer nuclear spin ($I=n/2$). For simplicity, here we take the $S=1/2,I=1/2$ system as an example. At zero magnetic field, the spin Hamiltonian of this electron-nuclei system is determined solely by the hyperfine interaction, and can be written as \cite{Cole1963}
\begin{equation}
H_{0} = A_{\perp}(S_{x}I_{x}+S_{y}I_{y})+A_{\parallel}S_{z}I_{z},
\label{P1Hamiltonian}
\end{equation}
where $A_{\perp}$ and $A_{\parallel}$ are the hyperfine constants. The eigenstates consist of one antisymmetric singlet $| S_{0} \rangle$ with $F=0$ and three symmetric triplet states $| T_{0} \rangle$ and $| T_{\pm 1} \rangle$ with $F=1$, where $\mathbf{F}=\mathbf{S}+\mathbf{I}$ is the total angular momentum. The corresponding eigenvalues are
\begin{equation}
\begin{split}
& \omega_{S_{0}} = -\frac{A_{\perp}}{2}-\frac{A_{\parallel}}{4}, \\
& \omega_{T_{0}} = \frac{A_{\perp}}{2}-\frac{A_{\parallel}}{4}, \\
& \omega_{T_{\pm1}} = \frac{A_{\parallel}}{4}.
\end{split}
\label{energies}
\end{equation}

\begin{figure}
\centering \includegraphics[width=1\columnwidth]{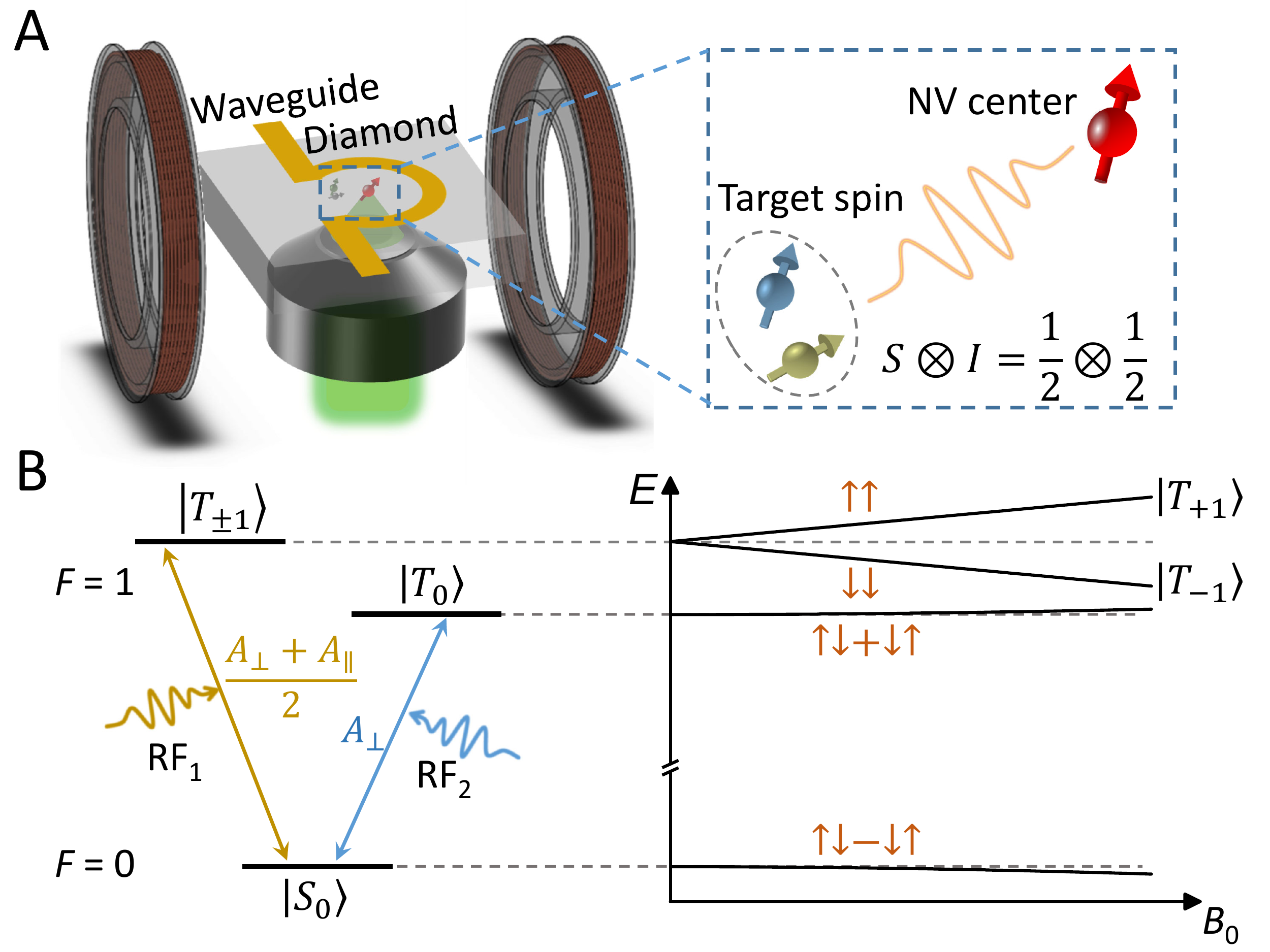} \protect\caption{
\textbf{Schematic representation of the NV center-based zero-field EPR spectrometer.}
(\textbf{A}) Geometry of the experimental setup. The sensor is a shallow NV center in diamond, which is observed by a confocal microscope with green-laser excitation and red-fluorescence collection. The $\Omega$-shape waveguide radiates microwave (MW) and radiofrequency (RF) to control the sensor and the target. The three-dimensional Helmholtz coils (only one of them is represented) are used to compensate residual magnetic fields. Inset gives our model, where the target consists of a spin-1/2 electron spin and a spin-1/2 nuclear spin.
(\textbf{B}) Energy levels of the target spin. The degeneracy of $| T_{\pm 1} \rangle$ is lifted by a magnetic field $B_0$ with a linearly dependent splitting, while $| S_{0} \rangle$ and $| T_{0} \rangle$ have zero first-order dependence on $B_0$. The $\text{ST}_{\pm 1}$ and $\text{ST}_{0}$ transitions can be driven by perpendicular RF$_1$ and parallel RF$_2$ pulses, respectively (section~S1). Up/down arrows denote the high-field spin-up/down states.
}
\label{setup}
\end{figure}

In the presence of magnetic noise $\delta \mathbf{b}$, the energy levels of the target spin will fluctuate, leading to line broadening. It can be described by a perturbation to the Hamiltonian:
\begin{equation}
\delta H = \sum_{j=x,y,z}\delta_j S_j,
\label{noise}
\end{equation}
where $\delta_j = \gamma_{\text{e}}\cdot \delta b_j$, $\gamma_{\text{e}}$ is the gyromagnetic ratio of the electron spin. Here we ignore the Zeeman terms of the nuclear spin because of the nearly three-orders smaller gyromagnetic ratio than electron spins. According to the perturbation theory, the energy level shifts can be simplified as (section~S1)
\begin{equation}
\begin{split}
& \delta\omega_{S_{0}} \approx -\frac{{\delta_x}^2+{\delta_y}^2}{2(A_{\parallel}+A_{\perp})}-\frac{{\delta_z}^2}{4A_{\perp}}, \\
& \delta\omega_{T_{0}} \approx -\frac{{\delta_x}^2+{\delta_y}^2}{2(A_{\parallel}-A_{\perp})}+\frac{{\delta_z}^2}{4A_{\perp}}, \\
& \delta\omega_{T_{\pm1}} \approx \pm\frac{\delta_z}{2}.
\end{split}
\label{energyshift}
\end{equation}
It shows that the $| S_{0} \rangle$ and $| T_{0} \rangle$ states have zero first-order dependence on magnetic field, and the frequency fluctuation of the $| S_{0} \rangle \rightarrow | T_{0} \rangle$ (denoted as $\text{ST}_0$ hereinafter) transition is reduced to $\sim \delta^2/A$ (Fig.~1B). Therefore, a line narrowing phenomenon will appear.

Such a narrowed $\text{ST}_0$ spectrum is challenging to observe by the previous microwave power-sweeping method \cite{Kong2018}, as it requires extreme power stability of the entire microwave circuits. The commonly used double electron-electron resonance (DEER) method \cite{Grinolds2013,Shi2015,Shi2018,Schlipf2017,Grotz2011,Mamin2012} is also not suitable, because the interrogation time is limited by the coherence time $T_2$ of the NV center. Inspired by the correlation spectroscopy of nuclear spins \cite{Laraoui2013}, we develop a modified correlation detection protocol for zero-field EPR spectroscopy. Then the sensor's lifetime can be increased to the spin-locking relaxation time $T_{1\rho}$, which is usually much longer than $T_2$ for shallow NV centers \cite{Kong2018,Rosskopf2014}.

As shown in Fig.~2A, our pulse protocol consists of two zero-field DEER sequences (see Materials and Methods) separated by a spin-locking sequence. After initialization of the NV center in $|0\rangle$ state by a 532 nm laser excitation, a resonant microwave $\pi$ pulse creates a superposition state $(|1\rangle+|-1\rangle)/\sqrt{2}$, which evolves during the first DEER period with an accumulated phase $\varphi_1$, and thus becomes $(e^{\text{i}\varphi_1}|1\rangle+e^{-\text{i}\varphi_1}|-1\rangle)/\sqrt{2}$. We rewrite it as $\cos{\varphi_1}|\psi_+\rangle +\text{i}\sin{\varphi_1}|\psi_-\rangle$, where $|\psi_{\pm}\rangle = (|1\rangle \pm |-1\rangle)/\sqrt{2}$. During the spin-locking period, both the $|\psi_{+}\rangle$ and $|\psi_{-}\rangle$ states can be locked by the continuous driving field, but the coherence between them vanishes, which can be regarded as an ensemble projection measurement (see Materials and Methods). In this period, we can perform any manipulations on the target spin, which determines the accumulated phase $\varphi_2$ during the second DEER period. If the NV state is projected to $|\psi_{+}\rangle$ (or $|\psi_{-}\rangle$) with probability of $\cos^2{\varphi_1}$ (or $\sin^2{\varphi_1}$) after the spin-locking period, the final state will be $\cos{\varphi_2}|0\rangle+\text{i}\sin{\varphi_2}|\psi_-\rangle$ (or $\sin{\varphi_2}|0\rangle-\text{i}\cos{\varphi_2}|\psi_-\rangle$) with the population of $|0\rangle$ read out by the photoluminescence (PL) rate of the NV center. The resulting correlation signal is given by
\begin{equation}
S_{\text{corr}} = \frac{1}{2}[1+\langle \cos{2\varphi_1}\cos{2\varphi_2} \rangle],
\label{Correlation Signal}
\end{equation}
where the brackets denote statistical average.

\begin{figure}
\centering \includegraphics[width=1\columnwidth]{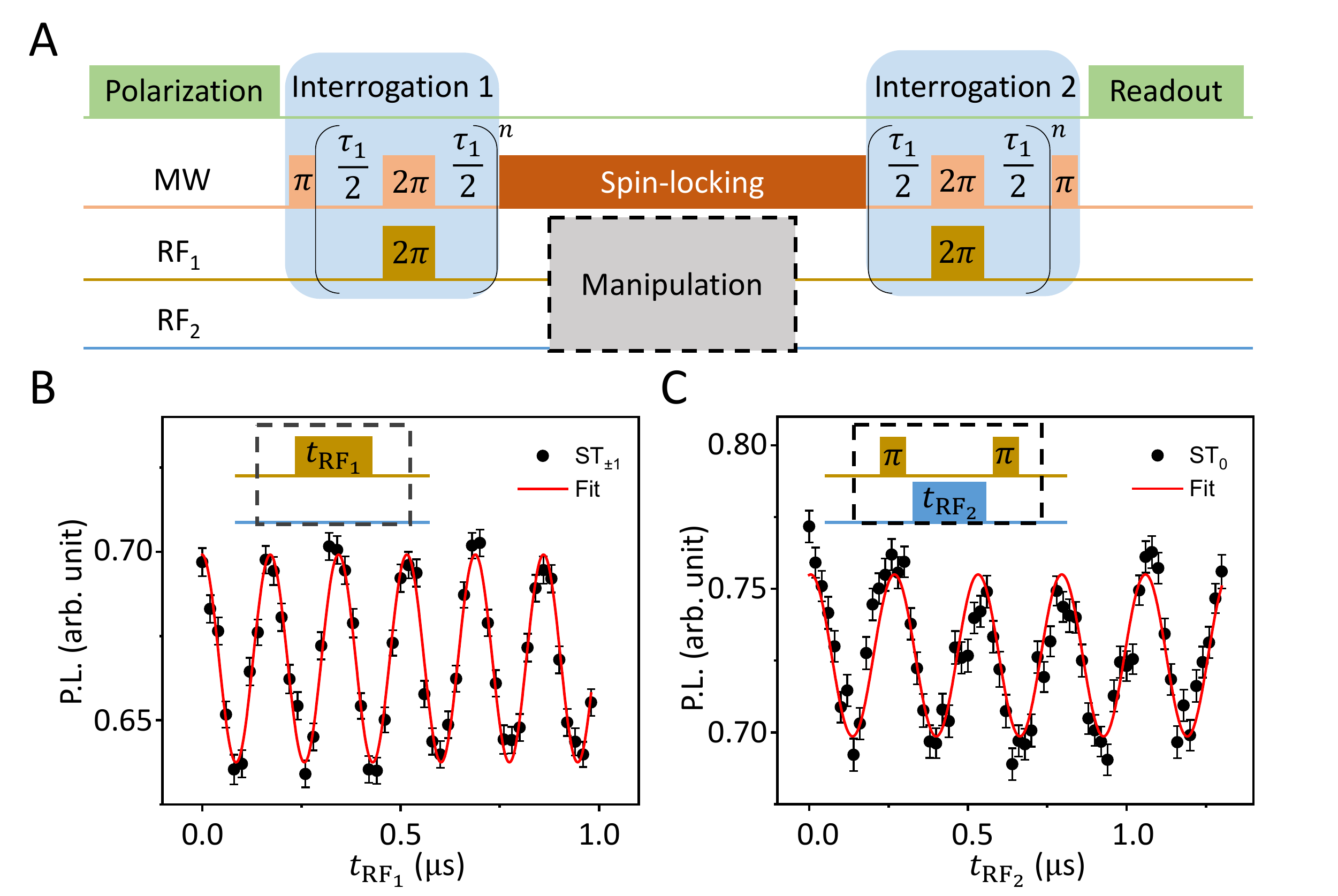} \protect\caption{
\textbf{Correlation protocol for EPR measurements.}
(\textbf{A}) Schematics of the pulse sequence. The interrogations are two DEER sequences ($n=2$) for detections of the target spin state. The detected phase signal is stored on the populations of NV dressed states and protected by spin locking. The correlation signal depends on the manipulations on the target spin, which is denoted by the black dash box.
(\textbf{B})(\textbf{C}) Rabi oscillations for the $\text{ST}_{\pm 1}$ (\textbf{B}) and $\text{ST}_{0}$ (\textbf{C}) transitions. Insets give the corresponding manipulations on P1 centers. The spin-locking time is fixed to 10 $\mu$s. The red lines are sine fittings. Error bars indicate s.e.m.
}
\label{Detection}
\end{figure}

We perform the experiments on a coupled system of NV and P1 centers at zero magnetic field. We compensate the residual magnetic field to $\sim$ 0.01 G by three-dimensional Helmholtz coils (see Materials and Methods and fig.~S1). P1 center is another kind of defect in diamond, consisting of only a substitutional nitrogen atom \cite{Hall2016,Kong2018}. The $^{15}$N P1 center is a $S=1/2$, $I=1/2$ system, making it an ideal candidate for demonstrations of the high-resolution zero-field EPR spectroscopy. Both the NV and P1 centers are created by implantation of $^{15}$N$_{2}^{+}$ ions into a bulk diamond. With proper energy and dose (see Materials and Methods), the created NV centers can be well resolved by a home-built confocal microscope. For some NV centers, one can find a single adjacent P1 center by performing the zero-field DEER measurements (see section~S2 and fig.~S2). The hyperfine constants of $^{15}$N P1 center are $A_{\perp}=114$ MHz and $A_{\parallel}=160$ MHz \cite{Lasher1959}, and thus the expected transition frequencies of $\text{ST}_{0}$ and $\text{ST}_{\pm 1}$ transitions are 114 MHz and 137 MHz, respectively.

The P1 center can be fully controlled by resonant radiofrequency (RF) pulses and reliably read out by the correlation detection protocol. Considering the orientation of P1 center jumps between the four kinds of N-C bonds (fig.~S1B) due to the Jahn-Teller effect \cite{Davies1981}, we adjust the RF direction to the vertical direction of the diamond surface by moving the P1 center to the central area of the $\Omega$-shape waveguide, so that the effective control-field strengths for the different orientated P1 centers are the same in order to simplify the control. Figure~2B shows the Rabi oscillation between $| S_{0} \rangle$ and $| T_{\pm1} \rangle$, by varying the RF pulse length during the spin-locking period. The spin-locking power, denoted by the corresponding Rabi frequency of the NV center, is 30 MHz, resulting in $T_{1\rho}\sim$ 150 $\mu$s. As a comparison, $T_2$ is just 16 $\mu$s (fig.~S3). This interrogation time can be further improved to $T_{1}$ ($\sim$ ms), although with compromise of signal contrast (see Materials and Methods). As described above, the $\text{ST}_{0}$ transition is insensitive to the magnetic field, and thus also insensitive to the coupling with the NV center. To observe this mute transition, we use $| T_{\pm1} \rangle$ as auxiliary states. As shown in Fig.~2C, we can also observe the Rabi oscillation between $| S_{0} \rangle$ and $| T_{0} \rangle$. The detailed calculations of these correlation Rabi measurements are in section~S3.

To obtain the EPR spectra, we perform the Ramsey experiments on the P1 center. Specifically, for the $\text{ST}_{\pm 1}$ transition, the manipulations during spin-locking period are two resonant $\pi/2$ RF pulses with varying separations $t$, as illustrated in Fig.~3A. Here the RF pulses remain unchanged during the variation of $t$ to avoid any spurious effect on the NV spin. So the phase difference between them varies proportionally to $t$, resulting in oscillation signals (Fig.~3A). The Fourier transformation gives the resonance spectrum. The decay of the signal remarks the decoherence process, resulting in line broadening in the frequency domain. Similarly, the Ramsey experiment of the $\text{ST}_{0}$ transition can also be performed with the assistance of the $\text{ST}_{\pm 1}$ transition, which shows a much slower decay. The detailed calculations of these correlation Ramsey measurements are in section~S4. Figure~3B gives the Fourier transformation spectra, and clearly shows the improvement of the spectral resolution. The Gaussian fittings give the linewidths (quoted as the full-width at half-maximum, FWHM) of the spectra, the minimums of which are $230\pm20$ kHz and $8.6\pm0.4$ kHz for the $\text{ST}_{\pm 1}$ and $\text{ST}_{0}$ transitions, respectively. The latter has ($27\pm3$)-fold improvement. If the line broadening is solely induced by magnetic noises, the estimated improvement should be $>$ 130 (see Materials and Methods). This deviation suggests that other decoherence sources emerge when the magnetic noise is suppressed, such as electric or strain field noises \cite{Kim2015}.

\begin{figure}
\centering \includegraphics[width=1\columnwidth]{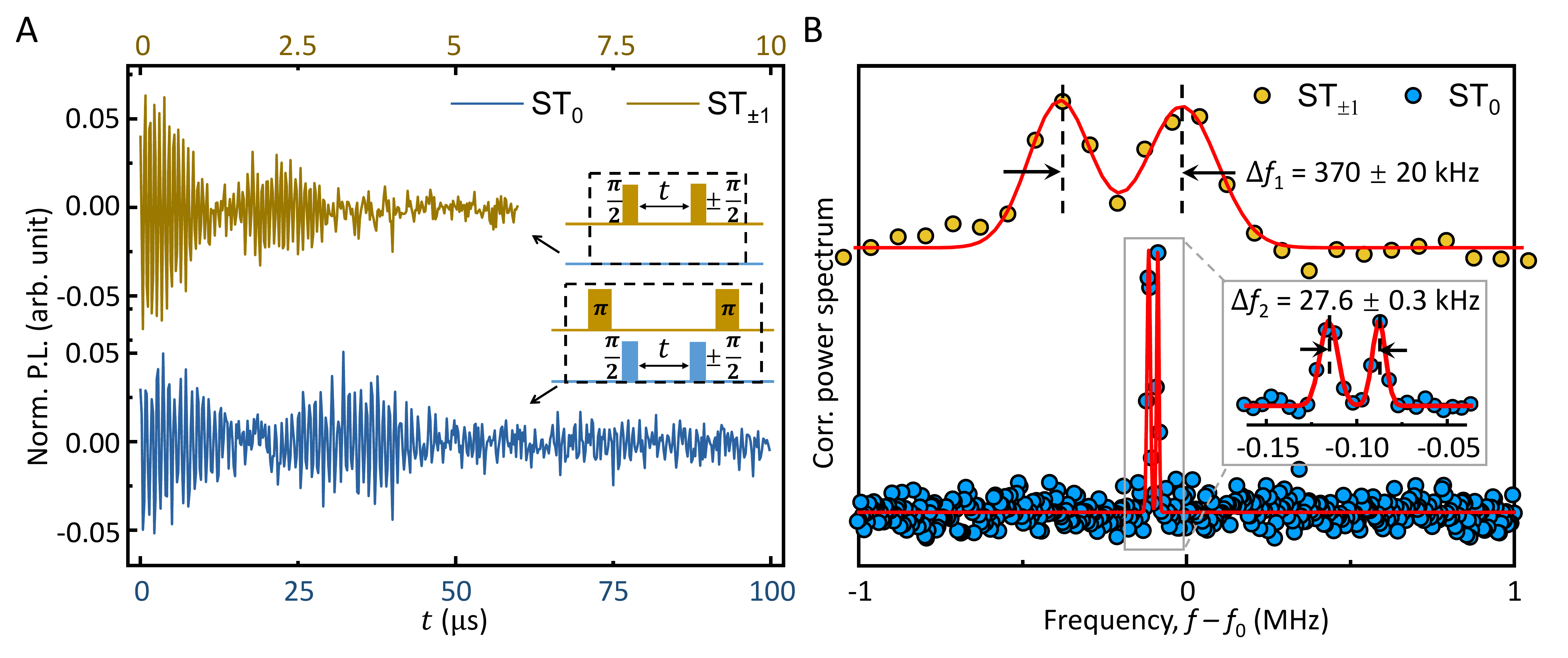} \protect\caption{
\textbf{High-resolution EPR spectroscopy of single P1 centers.}
(\textbf{A}) Correlation signals of Ramsey experiments for the $\text{ST}_{\pm 1}$ (upper) and $\text{ST}_{0}$ (down) transitions. Insets give the corresponding manipulations on P1 centers. Each experiment consists of two measurements, $\pi/2 \rightarrow -\pi/2$ and $\pi/2 \rightarrow \pi/2$, and the differential signal is presented. Note the upper data has a $10\times$ magnification in horizontal axis, and still shows faster decay. All the signals are undersampled, and the actual frequencies can be recovered with the prior knowledge of the rough resonance frequencies.
(\textbf{B}) Fourier transformations of the time-domain data. The frequencies are relative to 137 MHz ($\text{ST}_{\pm 1}$) and 114 MHz ($\text{ST}_{0}$). The points are experimental results while the solid line are two-Gaussian fittings. The fitting FWHM is 230$\pm$20 kHz (left) and 260$\pm$20 kHz (right) for the $\text{ST}_{\pm 1}$ spectra, and 11.6$\pm$0.6 kHz (left) and 8.6$\pm$0.4 kHz (right) for the $\text{ST}_{0}$ spectra.
}
\label{spectra}
\end{figure}

Both of the spectra show clear line splitting, but the reasons are different. For the $\text{ST}_{\pm 1}$ transition, the line splitting of 370 kHz is induced by the coupling with a nearby $^{13}$C nuclear spin, which is widespread in natural-isotope diamonds. As a comparison, we repeat the measurements on a $^{12}$C isotopically purified diamond, where the line splitting indeed disappears, but remains in the $\text{ST}_{0}$ spectrum (fig.~S4). Actually, the $^{13}$C coupling will not induce obvious splitting for the $\text{ST}_{0}$ transition, because of the quadratic dependence on the magnetic field according to Eq.~\ref{energyshift}. A probable reason for the $\text{ST}_{0}$ line splitting is the existence of local electric or strain fields\cite{Mittiga2018}. Similar with donor electron spins in silicon \cite{Bradbury2006,Wolfowicz2013}, the hyperfine coupling of P1 centers should also depend on electric or strain fields. Given that the hyperfine coupling here is anisotropic with jumping orientations, up to four $\text{ST}_{0}$ lines can emerge in a local static electric or strain field. The observation of only two lines suggests the field is along a symmetric direction, probably perpendicular to the surface, as reported in a recent paper \cite{Broadway2018}. Different from the magnetic field, the electric or strain field induces the same-order frequency shifts of the $\text{ST}_{0}$ and $\text{ST}_{\pm 1}$ transitions, because both them are linearly dependent on the hyperfine constants. Therefore, such weak couplings cannot be resolved by the $\text{ST}_{\pm 1}$ spectrum or previous non zero-field measurements.

To further investigate the spectral figure, we apply a small magnetic field $B$ and observe the spectral variance. As shown in Fig.~4A, further splittings appear with increasing $B$, which suggests that the $\text{ST}_{\pm 1}$ line splitting at zero field indeed comes from the coupling with a two-level system (i.e. $^{13}$C nuclear spin) rather than a local static magnetic field. On the other hand, the line shape of the $\text{ST}_{0}$ transition remains unchanged, despite the overall small shifts and line broadening (Fig.~4B). The line splitting is almost constant versus $B$, but varies for different P1 centers (fig.~S4), which suggests different local electric or strain environments \cite{Mittiga2018}. The resonance frequencies clearly show the linear and quadratic dependence on $B$ for the $\text{ST}_{\pm 1}$ and $\text{ST}_{0}$ transitions, respectively, which is well predicted by Eq.~\ref{energyshift}. The almost changeless peak positions in Fig.~4B also reveal the line narrowing nature, which is degrading with increasing $B$ due to the growing fluctuation of $(\gamma_{\text{e}}B+\delta)^2$.

\begin{figure}
\centering \includegraphics[width=1\columnwidth]{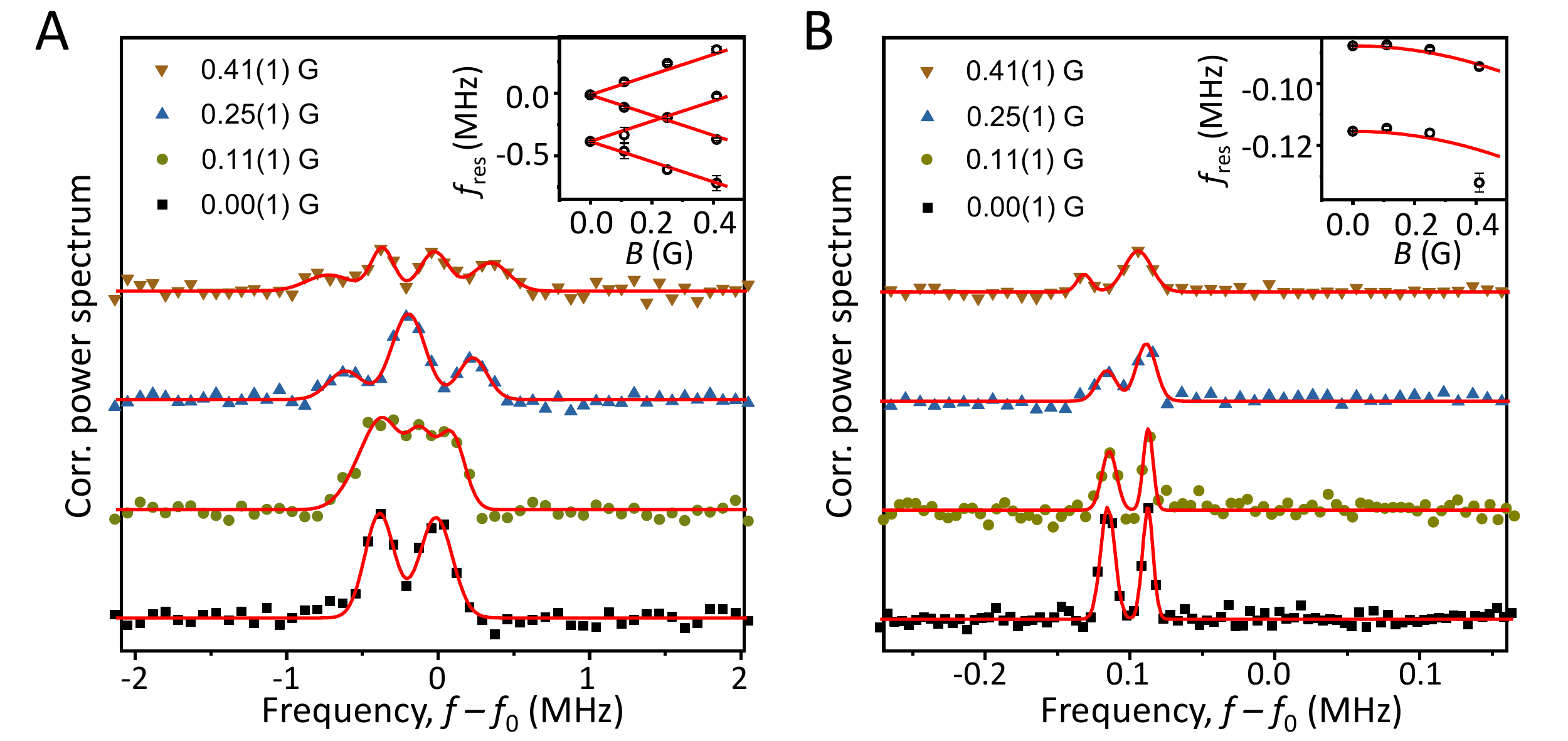} \protect\caption{
\textbf{Magnetic-field dependence of the line shape.}
The magnetic field $B$ is applied along the vertical direction of the diamond surface, which is a symmetric direction of the four P1 axis. All the points are experimental data, fitted with $m$-Gaussian ($m=2-4$) functions (solid lines). Insets give the fitting peak positions ($f_{\text{res}}$) versus magnetic fields with error bars defined by the fitting errors.
(\textbf{A}) $\text{ST}_{\pm 1}$ spectrum. Each peak at zero field splits independently with increasing $B$, and the expected splitting (solid lines in the inset) is calculated according to Eq.~\ref{energyshift}.
(\textbf{B}) $\text{ST}_{0}$ spectrum. The two-peak pattern remains unchange with increasing $B$, despite overall peak shifts and line broadening. Each group of peak positions (solid lines in the inset) are calculated according to Eq.~\ref{energyshift}.
}
\label{Bdepedence}
\end{figure}

In conclusion, we have presented a method for measuring EPR spectra with high spectral resolution, via employing a magnetic noise-insensitive singlet-triplet transition at zero magnetic field. The experiment demonstrates the EPR linewidth of P1 centers could be reduced by over one order of amplitude to several kilohertz, and then some weak non-magnetic effects can be discovered. Our results show that, as a sensor for EPR spectroscopy, the NV center has the potential to simultaneously achieve high spatial and high spectral resolution.

The P1 centers in diamond are ideal systems for demonstrations of this technique, while practical applications require target spins external to the diamond surface. Although this line narrowing phenomenon also exists in many other paramagnetic materials, such as $^{15}$N nitroxide spin labels \cite{Shi2018} and some transition ions \cite{Erickson1966,Steven1984}, the observations with NV centers are more challenging. First, fast rotations of the target spins will cancel their dipolar couplings with the NV sensor in the absence of external magnetic field, making the detections impossible. So the sample should be solid or quasi solid. Fortunately, the zero-field powder spectra do not have the inhomogeneous line broadening issue \cite{Cole1963,Erickson1966,Steven1984,Kong2018}. Secondly, the spectral improvement will degrade with increasing magnetic noise strength as $\sim \delta/A$, and will vanish when the noise strength $\delta$ is comparable to or even larger than the hyperfine constant $A$. It means the coherence properties of the target spins should not be too worse relative to $1/A$. This may be a problem for the transition ions, whose coherence times are usually much worse than free-radical spin labels. Nevertheless, it is recently reported that \cite{Zadrozny2015}, with proper synthetic design, the coherence times of some transition metal complexes can even approach millisecond. Thirdly, the current measurement is still time consuming. The widely used nitroxide spin labels may be bleached during the strong laser illumination \cite{Schlipf2017,Shi2018}. A possible solution is improving the detection efficiency by implanting spin-to-charge readout technique \cite{Shields2015}, or extending the lifetimes of spin labels by lowering the temperature and moving to a vacuum environment \cite{Schlipf2017}.

If successfully generalizing to external paramagnetic targets, this technique will benefit many EPR studies with significantly improved precision. For example, zero-field EPR spectroscopy is a powerful tool to unambiguously extract the hyperfine constants \cite{Kong2018}, which can reflect the local polarity profiles \cite{Kurad2003}. By improving spectral resolution, more detailed micro-environment information can be explored. In addition, such magnetic field-insensitive singlet and triplet states can serve as a narrow-band filter, which filters out magnetic fields of all frequencies except the one resonant with the transition. If considering dual spin targets in the presence of spin bath, all the target-environment spin dipolar couplings are killed, but the target-target coupling survives. Thus the latter can be precisely detected by deploying our zero-field EPR technique, which offers interesting avenues for enhanced accuracy and detection range in distance measurement of spin-modified biomolecules \cite{Jeschke2012}.

\section*{MATERIALS AND METHODS}
\subsection*{Experimental setup and diamond samples}
The optical part of our setup is a home-built confocal microscopy, with a diode laser (CNI MGL-III-532) used for illumination and an avalanche photodiode (Perkin Elmer SPCM-AQRH-14) used for photons collection. All the microwave and radiofrequency pulses are generated by an arbitrary waveform generator (Keysight M8190a) and amplified by two amplifiers accordingly (Mini-circuits ZHL-16W-43+ and LZY-22+). The $\Omega$-shape waveguide has a diameter of 50 $\mu$m with radiation field perpendicular to the plane in the central area.

Two diamond samples are involved, denoted as diamond A and B. Both of them are obtained commercially, 100-oriented, electronic-grade, and implanted with $^{15}$N$_{2}^{+}$ ions. Diamond A is implanted with energy of 30 keV and dose of $3\times10^{10}$ $\text{cm}^{-2}$. We fabricated nanopillars on diamond A to enhance the photon-collection efficiency \cite{Hausmann2011}. Diamond B has an extra growth layer with 99.9\% $^{12}$C isotopic purity. It is implanted with energy of 5 keV and dose of $1.7\times10^{10}$ $\text{cm}^{-2}$. The simulated ion straggling ($\sim$ 7 nm for diamond A, and $\sim$ 4 nm for diamond B) is much smaller than the mean spacing of ions ($\sim$ 60 nm for diamond A, and $\sim$ 80 nm for diamond B), so the detected signal is dominated by a single P1 center arising from the adjacent nitrogen ion. The data in fig.~S4 are measured on diamond B, and all the other data are measured on diamond A.

\subsection*{Compensation of residual magnetic field}

The NV center can detect magnetic fields via optically detected magnetic resonance (ODMR) spectrum. In small magnetic field ($\gamma_{\text{e}}B \ll D$), the line splitting is proportional to the component parallel to the N-V axis. However, the smallest resolvable magnetic field is limited by the ODMR linewidth. To reliably compensate the residual magnetic field, we take advantage of the symmetric nature of the ODMR spectrum versus magnetic fields. As shown in fig.~S1A, with increasing currents applied to the coils, the ODMR linewidth first decreases and then increases. The symmetric center point remarks the zero axial magnetic field. By utilizing three differently orientated NV centers, the residual vector magnetic field $\mathbf{B}$ can be totally compensated. Figure~S1B gives the schematics of the three NV centers, among which NV A/B and C are orientated perpendicular to the X and Y axis, respectively. We perform the compensation as followings: (i) Compensation of $B_z$. The ODMR linewidths of NV A and B are measured with sweeping Z-coil currents, and two symmetric center points $I_{Z,A}$ and $I_{Z,B}$ are recorded (fig.~S1C). Then the Z-coil compensation current is $I_{Z}=(I_{Z,A}+I_{Z,B})/2$ due to the symmetry of NV A and B with respect to Z axis. (ii) Compensation of $B_y$. Since the NV A and B are insensitive to $B_x$, the compensation of $B_y$ is straightforward. As shown in fig.~S1D, the symmetric center point is just the Y-coil compensation current. (iii) Compensation of $B_x$. Similar with (ii), the symmetric center point of NV C is just the X-coil compensation current (fig.~S1E). The maximum fit error of the symmetric center point is 1 mA, while the stability of the power supplies is $\sim$ 4 mA. The magnetic field linearly depends on the current with coefficient of 2.8 G/A. Therefore, the magnetic field after compensation is estimated to be $\sim$ 0.01 G.

\subsection*{Zero-field DEER}

For NV center-based EPR, the double electron-electron resonance (DEER) sequence is widely used for the detection of electron spins \cite{Grinolds2013,Shi2015,Shi2018,Grotz2011,Mamin2012}, where the NV center itself is also a electron spin ($S=1$). Due to the degeneracy of $|\pm 1\rangle$ at zero magnetic field, the zero-field DEER sequence is slightly different (fig.~S2). Specifically, a $\pi$ pulse can flip the NV state from $|0\rangle$ to a superposition state $(|1\rangle + |-1\rangle)/\sqrt{2}$, which serves as an interferometer, and a $2\pi$ pulse can switch $|1\rangle$ and $|-1\rangle$, which is a decoupling operation. During the evolution, the superposition state accumulates a phase $\phi = a\cdot \tau$, where $a$ is the dipolar coupling strength, and becomes $(e^{\text{i}\phi}|+1\rangle+e^{-\text{i}\phi}|-1\rangle)/\sqrt{2}$. Finally, a second $\pi$ pulse reverses this state back to $|0\rangle$ with the population $\cos^{2}{\phi}$ read out by the photoluminescence rate.

The dipole-dipole coupling between the sensor and the target can be approached by
\begin{equation}
H_{\text{dd}} \approx C(\mathbf{n}_{\text{NV}},\mathbf{n}_{\text{tar}},\mathbf{r})S_{z}^{\text{NV}}S_{zz}^{\mathcal{T}},
\label{simple dd}
\end{equation}
where the coupling strength $C$ depends on the orientations of the NV center $\mathbf{n}_{\text{NV}}$, of the target spin $\mathbf{n}_{\text{tar}}$, and the spacing vector $\mathbf{r}$ between them. $S_{zz}^{\mathcal{T}}$ is a reduced spin operator of the target spin in the singlet-triplet basis $\{| T_{+1} \rangle, | S_{0} \rangle, | T_{0} \rangle, | T_{-1} \rangle\}$:
\begin{equation}
S_{zz}^{\mathcal{T}} =
\frac{1}{2}\begin{pmatrix}
1 &   &   &  \\
  & 0 &   &   \\
  &   & 0 &  \\
  &   &   & -1
\end{pmatrix}.
\label{SzzR}
\end{equation}
The detailed derivation process is in section~S1. Now one can directly see that the coupling will disappear if the target spin is in the $| S_{0} \rangle$ or $| T_{0} \rangle$ states, consisting with the magnetic field-insensitive nature. So the DEER measurement can not directly capture the signal of $\text{ST}_{0}$ transition. Otherwise, if the target spin is in the $| T_{\pm1} \rangle$ states, the coupling strength $a=\pm C/2$. Similar with the NV center, a $\pi$ RF pulse will flip the target spin from $| S_{0} \rangle$ to a superposition state of $| T_{\pm 1} \rangle$, while a $2\pi$ RF pulse will switch $| T_{+1} \rangle$ and $| T_{-1} \rangle$. The latter leads to a stronger equivalent coupling strength (fig.~S2), which is more favorable in the presence of decoherence. As calculated in section~S2, the zero-field DEER signal is
\begin{equation}
S(\tau) = \frac{3}{4}+\frac{1}{4}e^{-(\tau/T_{2,\text{NV}})^p}\cos{C\tau},
\label{DEER_signal2}
\end{equation}
where the decoherence of the NV center is described by a stretched exponential decay, and $p$ is in the range of $1-3$ determined by the dynamic of bath \cite{BarGill2012}.

\subsection*{Zero-field spin locking}
The Hamiltonian of the NV center at zero magnetic field is
\begin{equation}
H_0 = DS_{z}^{2},
\end{equation}
where $\mathbf{S} (S=1)$ is the NV electron spin operators, $D$ is the zero-field splitting. During the spin-locking period, we apply a phase modulated microwave of the form \cite{Cohen2017}
\begin{equation}
H_{1} = \Omega_1 \cos{[Dt+\frac{2\Omega_2}{\Omega_1}\sin{\Omega_1 t}]}S_{x},
\end{equation}
where $\Omega_1$ is the corresponding Rabi frequency, and $\Omega_2$ is the phase modulation strength. In this experiment, $\Omega_2=0.3\Omega_1$. By moving to the interaction picture, the Hamiltonian becomes
\begin{equation}
\begin{split}
H_{\text{I}} &= e^{\text{i}f(t)S_{z}^{2}}(H_0+H_1)e^{-\text{i}f(t)S_{z}^{2}}-f^{'}(t)S_{z}^{2} \\
& = \frac{\Omega_1}{2}S_{x}-2\Omega_2 \cos{\Omega_1 t}S_{z}^{2},
\end{split}
\label{xphase}
\end{equation}
where $f(t)=Dt+(2\Omega_2/\Omega_1)\sin{\Omega_1 t}$, and we ignore the high-frequency items. This is just the common Rabi model. Moving again to the second interaction picture and ignoring the high-frequency items, we can write
\begin{equation}
\begin{split}
H_{\text{II}} &= e^{\text{i}\frac{\Omega_1 t}{2}S_{x}}H_{\text{I}}e^{-\text{i}\frac{\Omega_1 t}{2}S_{x}}-\frac{\Omega_1}{2}S_{x} \\
& = -\frac{\Omega_2}{2}(S_{z}^{2}-S_{y}^{2}).
\end{split}
\label{xphase2}
\end{equation}
For an initial state $|\psi(0)\rangle$, the evolution of the NV state can be written as
\begin{equation}
|\psi(t)\rangle = e^{-\text{i}f(t)S_{z}^{2}}e^{-\text{i}\frac{\Omega_1}{2}S_{x}t}e^{-iH_{\text{II}}t}|\psi(0)\rangle.
\end{equation}
By choosing an integral-period evolution time, i.e., $t=k\cdot 4\pi/\Omega_1, k=1,2,\cdots$, the evolution is simplified to
\begin{equation}\label{lockingevo}
|\psi(t)\rangle = e^{-\text{i}H_{\text{II}}t}|\psi(0)\rangle,
\end{equation}
where the phase item $\exp[-\text{i}DtS_{z}^{2}]$ is absorbed in the normal rotating reference frame. After the first DEER period, the NV state becomes $\cos{\phi_1}|\psi_+\rangle +\text{i}\sin{\phi_1}|\psi_-\rangle$. Note both $|\psi_{+}\rangle$ and $|\psi_{-}\rangle$ are eigenstates of $H_{\text{II}}$, so the spin-locking process is just a free induction decay process in the interaction picture. The populations of $|\psi_{\pm}\rangle$ are protected while the coherence between them is destroyed by the environmental noise. Therefore, the NV state becomes a mixed state of $|\psi_{+}\rangle$ and $|\psi_{-}\rangle$ after the spin-locking period.

\subsection*{Alternative protocol for correlation detection}

Alternatively, we can also avoid the spin-locking microwave in some further practical applications, such as sensing in a living cell, where long microwave pulses maybe potentially harmful to biological tissues. This correlation protocol consists of two DEER sequence separated by a free evolution (fig.~S5). Similar to the analysis in the main text, the NV state starts from $|0\rangle$, and becomes $\cos{\varphi_1}|0\rangle+\text{i}\sin{\varphi_1}|\psi_-\rangle$ after the first DEER period. During the free evolution, the coherence vanishes with NV state projecting to either $|0\rangle$ with probability of $\cos^2{\varphi_1}-1/2\sin^2{\varphi_1}$, or a totally mixed state $(|0\rangle\langle0|+|{1}\rangle\langle{1}|+|{-1}\rangle\langle{-1}|)/3$ with probability of $3/2\sin^2{\varphi_1}$. The former has a similar evolution as the first DEER period, and becomes $\cos{\varphi_2}|0\rangle+\text{i}\sin{\varphi_2}|\psi_-\rangle$ after the second DEER period, while the latter has an idle evolution. Therefore, the correlation signal is
\begin{equation}
  \begin{split}
  & S_{\text{corr}}(\varphi_1,\varphi_2) = \langle (\cos^2{\varphi_1}-\frac{1}{2}\sin^2{\varphi_1})\cos^2{\varphi_2}+\frac{1}{2}\sin^2{\varphi_1} \rangle \\
  & = \frac{1}{8}[3+\langle \cos{2\varphi_1} \rangle + \langle \cos{2\varphi_2} \rangle + 3\langle \cos{2\varphi_1}\cos{2\varphi_2} \rangle].
  \end{split}
\end{equation}
Here one can see that the signal contrast is reduced by 25\%, which is the cost of removing the spin-locking microwave. Nevertheless, the limitation of interrogation duration can be further released to $T_{1,\text{NV}}$.

\subsection*{Spectral linewidth analysis}

According to Eq.~\ref{energyshift}, the fluctuations of the transition frequencies can be approached by
\begin{equation}
\begin{split}
& \delta\omega_{\text{ST}_{\pm 1}} \approx \pm\frac{1}{2}\delta_z, \\
& \delta\omega_{\text{ST}_{0}} \approx -\frac{A_{\perp}}{A_{\parallel}^2-A_{\perp}^2}({\delta_x}^2+{\delta_y}^2)+\frac{1}{2A_{\perp}}{\delta_z}^2. \\
\end{split}
\label{fshift}
\end{equation}
For Ramsey measurements, the magnetic noise is dominated by the low-frequency component, which can be modeled as a quasi-static random variable characterized by a normal distribution
\begin{equation}
P(\delta=x)=\frac{1}{\sqrt{2\pi}\sigma}e^{-\frac{x^{2}}{2\sigma^{2}}},\label{Gauss noise}
\end{equation}
with standard deviation $\sigma$ used to characterize the noise amplitude. Accordingly, the energy levels will fluctuate and also can be modeled as some kinds of distributions, of which the standard deviations can be used to estimate the linewidth. According to Eq.~\ref{fshift}, the standard deviation of $\omega_{\text{ST}_{\pm 1}}$ can be directly written as
\begin{equation}
\sigma_{\text{ST}_{\pm 1}}=\frac{1}{2}\sigma.
\end{equation}
Considering isotropic magnetic noise, i.e., $\sigma_x = \sigma_y =\sigma_z = \sigma$, the standard deviation of $\omega_{\text{ST}_{0}}$ can be calculated as
\begin{equation}
\begin{split}
\sigma_{\text{ST}_{0}} & =\sqrt{\langle (\delta\omega_{\text{ST}_{0}})^2 \rangle - \langle \delta\omega_{\text{ST}_{0}} \rangle^2}\\
& =\sigma^2 \sqrt{\frac{4A_{\perp}^2}{(A_{\parallel}^2-A_{\perp}^2)^2}+\frac{1}{2A_{\perp}^2}},
\end{split}
\end{equation}
where $\langle\bullet\rangle = \int\int\int\bullet P(\delta_x)P(\delta_y)P(\delta_z)d\delta_x d\delta_y d\delta_z$. Thus, the spectral-resolution improvement is estimated to be
\begin{equation}
\chi \sim \frac{\sigma_{\text{ST}_{\pm 1}}}{\sigma_{\text{ST}_{0}}}= \left[\frac{64A_{\perp}^2}{(A_{\parallel}^2-A_{\perp}^2)^2}+\frac{8}{A_{\perp}^2}\right]^{-\frac{1}{2}} \cdot
\left(\sigma_{\text{ST}_{\pm 1}}\right)^{-1}.
\end{equation}
For the data in Fig.~3, the linewidth of the $\text{ST}_{\pm 1}$ spectrum is 230 kHz, which is defined by the FWHM. So $\sigma_{\text{ST}_{\pm 1}} = \text{FWHM}/\sqrt{8\ln 2} \approx 98$ kHz, and $\chi \sim$ 130. Actually, this improvement is underestimated, because the distribution of the $\text{ST}_{0}$ transition is not Gaussian but a sharper type (fig.~S6). The quadratic dependence on magnetic field is responsible to the asymmetric line shape. Both the disappearance of the asymmetric pattern and the degrading spectral-resolution improvement of the experimentally measured $\text{ST}_0$ spectrum suggest the existence of other decoherence resources. The magnetic noise can be increased by applying noise currents to the Helmholtz coils, and then the asymmetric pattern appears (fig.~S6).

\bibliographystyle{Science}

\begin{thebibliography}{10}

\bibitem{Borbat2001}
P.~P. Borbat, A.~J. Costa-Filho, K.~A. Earle, J.~K. Moscicki, J.~H. Freed,
  Electron spin resonance in studies of membranes and proteins.
\newblock {\it Science\/} {\bf 291}, 266--269 (2001).

\bibitem{Artzi2015}
Y.~Artzi, Y.~Twig, A.~Blank, Induction-detection electron spin resonance with
  spin sensitivity of a few tens of spins.
\newblock {\it Appl. Phys. Lett.\/} {\bf 106}, 084104 (2015).

\bibitem{Rugar2004}
D.~Rugar, R.~Budakian, H.~J. Mamin, B.~W. Chui, Single spin detection by
  magnetic resonance force microscopy.
\newblock {\it Nature\/} {\bf 430}, 329-332 (2004).

\bibitem{Baumann2015}
S.~Baumann, W.~Paul, T.~Choi, C.~P. Lutz, A.~Ardavan, A.~J. Heinrich, Electron
  paramagnetic resonance of individual atoms on a surface.
\newblock {\it Science\/} {\bf 350}, 417--420 (2015).

\bibitem{Grinolds2013}
M.~S. Grinolds, S.~Hong, P.~Maletinsky, L.~Luan, M.~D. Lukin, R.~L. Walsworth,
  A.~Yacoby, Nanoscale magnetic imaging of a single electron spin under ambient
  conditions.
\newblock {\it Nat. Phys.\/} {\bf 9}, 215-219 (2013).

\bibitem{Shi2015}
F.~Shi, Q.~Zhang, P.~Wang, H.~Sun, J.~Wang, X.~Rong, M.~Chen, C.~Ju,
  F.~Reinhard, H.~Chen, J.~Wrachtrup, J.~Wang, J.~Du, Single-protein spin
  resonance spectroscopy under ambient conditions.
\newblock {\it Science\/} {\bf 347}, 1135--1138 (2015).

\bibitem{Shi2018}
F.~Shi, F.~Kong, P.~Zhao, X.~Zhang, M.~Chen, S.~Chen, Q.~Zhang, M.~Wang, X.~Ye,
  Z.~Wang, Z.~Qin, X.~Rong, J.~Su, P.~Wang, P.~Z. Qin, J.~Du, Single-{DNA}
  electron spin resonance spectroscopy in aqueous solutions.
\newblock {\it Nat. Methods\/} {\bf 15}, 697-699 (2018).

\bibitem{Schlipf2017}
L.~Schlipf, T.~Oeckinghaus, K.~Xu, D.~B.~R. Dasari, A.~Zappe, F.~F.
  de~Oliveira, B.~Kern, M.~Azarkh, M.~Drescher, M.~Ternes, K.~Kern,
  J.~Wrachtrup, A.~Finkler, A molecular quantum spin network controlled by a
  single qubit.
\newblock {\it Sci. Adv.\/} {\bf 3} (2017).

\bibitem{Grotz2011}
B.~Grotz, J.~Beck, P.~Neumann, B.~Naydenov, R.~Reuter, F.~Reinhard, F.~Jelezko,
  J.~Wrachtrup, D.~Schweinfurth, B.~Sarkar, P.~Hemmer, Sensing external spins
  with nitrogen-vacancy diamond.
\newblock {\it New J. Phys.\/} {\bf 13}, 055004 (2011).

\bibitem{Mamin2012}
H.~J. Mamin, M.~H. Sherwood, D.~Rugar, Detecting external electron spins using
  nitrogen-vacancy centers.
\newblock {\it Phys. Rev. B\/} {\bf 86}, 195422 (2012).

\bibitem{Hall2016}
L.~T. Hall, P.~Kehayias, D.~A. Simpson, A.~Jarmola, A.~Stacey, D.~Budker,
  L.~C.~L. Hollenberg, Detection of nanoscale electron spin resonance spectra
  demonstrated using nitrogen-vacancy centre probes in diamond.
\newblock {\it Nat. Commun.\/} {\bf 7}, 10211 (2016).

\bibitem{Kong2018}
F.~Kong, P.~Zhao, X.~Ye, Z.~Wang, Z.~Qin, P.~Yu, J.~Su, F.~Shi, J.~Du,
  Nanoscale zero-field electron spin resonance spectroscopy.
\newblock {\it Nat. Commun.\/} {\bf 9}, 1563 (2018).

\bibitem{Martorana2014}
A.~Martorana, G.~Bellapadrona, A.~Feintuch, E.~Di~Gregorio, S.~Aime,
  D.~Goldfarb, Probing protein conformation in cells by {EPR} distance
  measurements using {Gd$^{3+}$} spin labeling.
\newblock {\it J. Am. Chem. Soc.\/} {\bf 136}, 13458-13465 (2014).

\bibitem{Kurad2003}
D.~Kurad, G.~Jeschke, D.~Marsh, Lipid membrane polarity profiles by high-field
  {EPR}.
\newblock {\it Biophys. J.\/} {\bf 85}, 1025-1033 (2003).

\bibitem{Jeschke2012}
G.~Jeschke, {DEER} distance measurements on proteins.
\newblock {\it Annu. Rev. Phys. Chem.\/} {\bf 63}, 419-446 (2012).

\bibitem{Laraoui2013}
A.~Laraoui, F.~Dolde, C.~Burk, F.~Reinhard, J.~Wrachtrup, C.~A. Meriles,
  High-resolution correlation spectroscopy of {$^{13}$C} spins near a
  nitrogen-vacancy centre in diamond.
\newblock {\it Nat. Commun.\/} {\bf 4}, 1651 (2013).

\bibitem{Zaiser2016}
S.~Zaiser, T.~Rendler, I.~Jakobi, T.~Wolf, S.-Y. Lee, S.~Wagner, V.~Bergholm,
  T.~Schulte-Herbr{\"u}ggen, P.~Neumann, J.~Wrachtrup, Enhancing quantum
  sensing sensitivity by a quantum memory.
\newblock {\it Nat. Commun.\/} {\bf 7}, 12279 (2016).

\bibitem{Sarkar2010}
R.~Sarkar, P.~Ahuja, P.~R. Vasos, G.~Bodenhausen, Long-lived coherences for
  homogeneous line narrowing in spectroscopy.
\newblock {\it Phys. Rev. Lett.\/} {\bf 104}, 053001 (2010).

\bibitem{Bollinger1985}
J.~J. Bollinger, J.~D. Prestage, W.~M. Itano, D.~J. Wineland,
  Laser-cooled-atomic frequency standard.
\newblock {\it Phys. Rev. Lett.\/} {\bf 54}, 1000--1003 (1985).

\bibitem{Harty2014}
T.~P. Harty, D.~T.~C. Allcock, C.~J. Ballance, L.~Guidoni, H.~A. Janacek, N.~M.
  Linke, D.~N. Stacey, D.~M. Lucas, High-fidelity preparation, gates, memory,
  and readout of a trapped-ion quantum bit.
\newblock {\it Phys. Rev. Lett.\/} {\bf 113}, 220501 (2014).

\bibitem{Wolfowicz2013}
G.~Wolfowicz, A.~M. Tyryshkin, R.~E. George, H.~Riemann, N.~V. Abrosimov,
  P.~Becker, H.-J. Pohl, M.~L.~W. Thewalt, S.~A. Lyon, J.~J.~L. Morton, Atomic
  clock transitions in silicon-based spin qubits.
\newblock {\it Nat. Nanotechnol.\/} {\bf 8}, 561-564 (2013).

\bibitem{Morse2018}
K.~J. Morse, P.~Dluhy, J.~Huber, J.~Z. Salvail, K.~Saeedi, H.~Riemann, N.~V.
  Abrosimov, P.~Becker, H.-J. Pohl, S.~Simmons, M.~L.~W. Thewalt, Zero-field
  optical magnetic resonance study of phosphorus donors in 28-silicon.
\newblock {\it Phys. Rev. B\/} {\bf 97}, 115205 (2018).

\bibitem{Emondts2014}
M.~Emondts, M.~P. Ledbetter, S.~Pustelny, T.~Theis, B.~Patton, J.~W. Blanchard,
  M.~C. Butler, D.~Budker, A.~Pines, Long-lived heteronuclear spin-singlet
  states in liquids at a zero magnetic field.
\newblock {\it Phys. Rev. Lett.\/} {\bf 112}, 077601 (2014).

\bibitem{Cole1963}
T.~Cole, T.~Kushida, H.~C. Heller, Zero field electron magnetic resonance in
  some inorganic and organic radicals.
\newblock {\it J. Chem. Phys.\/} {\bf 38}, 2915-2924 (1963).

\bibitem{Erickson1966}
L.~E. Erickson, Electron-paramagnetic-resonance absorption by trivalent
  neodymium ions in single crystals of lanthanum trichloride and lanthanum
  ethyl sulphate in zero magnetic field.
\newblock {\it Phys. Rev.\/} {\bf 143}, 295--303 (1966).

\bibitem{Steven1984}
S.~J. Strach, R.~Bramley, {EPR} of the vanadyl ion in {Tutton} salts at zero
  magnetic field.
\newblock {\it Chem. Phys. Lett.\/} {\bf 109}, 363-367 (1984).

\bibitem{Bramley1983}
R.~Bramley, S.~J. Strach, Electron paramagnetic resonance spectroscopy at zero
  magnetic field.
\newblock {\it Chem. Rev.\/} {\bf 83}, 49-82 (1983).

\bibitem{Zheng2019}
H.~Zheng, J.~Xu, G.~Z. Iwata, T.~Lenz, J.~Michl, B.~Yavkin, K.~Nakamura,
  H.~Sumiya, T.~Ohshima, J.~Isoya, J.~Wrachtrup, A.~Wickenbrock, D.~Budker,
  Zero-field magnetometry based on nitrogen-vacancy ensembles in diamond.
\newblock {\it Phys. Rev. Applied\/} {\bf 11}, 064068 (2019).

\bibitem{Rosskopf2014}
T.~Rosskopf, A.~Dussaux, K.~Ohashi, M.~Loretz, R.~Schirhagl, H.~Watanabe,
  S.~Shikata, K.~M. Itoh, C.~L. Degen, Investigation of surface magnetic noise
  by shallow spins in diamond.
\newblock {\it Phys. Rev. Lett.\/} {\bf 112}, 147602 (2014).

\bibitem{Lasher1959}
W.~V. Smith, P.~P. Sorokin, I.~L. Gelles, G.~J. Lasher, Electron-spin resonance
  of nitrogen donors in diamond.
\newblock {\it Phys. Rev.\/} {\bf 115}, 1546--1552 (1959).

\bibitem{Davies1981}
G.~Davies, The {J}ahn-{T}eller effect and vibronic coupling at deep levels in
  diamond.
\newblock {\it Rep. Prog. Phys.\/} {\bf 44}, 787--830 (1981).

\bibitem{Kim2015}
M.~Kim, H.~J. Mamin, M.~H. Sherwood, K.~Ohno, D.~D. Awschalom, D.~Rugar,
  Decoherence of near-surface nitrogen-vacancy centers due to electric field
  noise.
\newblock {\it Phys. Rev. Lett.\/} {\bf 115}, 087602 (2015).

\bibitem{Mittiga2018}
T.~Mittiga, S.~Hsieh, C.~Zu, B.~Kobrin, F.~Machado, P.~Bhattacharyya, N.~Z.
  Rui, A.~Jarmola, S.~Choi, D.~Budker, N.~Y. Yao, Imaging the local charge
  environment of nitrogen-vacancy centers in diamond.
\newblock {\it Phys. Rev. Lett.\/} {\bf 121}, 246402 (2018).

\bibitem{Bradbury2006}
F.~R. Bradbury, A.~M. Tyryshkin, G.~Sabouret, J.~Bokor, T.~Schenkel, S.~A.
  Lyon, Stark tuning of donor electron spins in silicon.
\newblock {\it Phys. Rev. Lett.\/} {\bf 97}, 176404 (2006).

\bibitem{Broadway2018}
D.~A. Broadway, N.~Dontschuk, A.~Tsai, S.~E. Lillie, C.~T.-K. Lew, J.~C.
  McCallum, B.~C. Johnson, M.~W. Doherty, A.~Stacey, L.~C.~L. Hollenberg, J.-P.
  Tetienne, Spatial mapping of band bending in semiconductor devices using in
  situ quantum sensors.
\newblock {\it Nat. Electron.\/} {\bf 1}, 502--507 (2018).

\bibitem{Zadrozny2015}
J.~M. Zadrozny, J.~Niklas, O.~G. Poluektov, D.~E. Freedman, Millisecond
  coherence time in a tunable molecular electronic spin qubit.
\newblock {\it ACS Cent. Sci.\/} {\bf 1}, 488-492 (2015).

\bibitem{Shields2015}
B.~J. Shields, Q.~P. Unterreithmeier, N.~P. de~Leon, H.~Park, M.~D. Lukin,
  Efficient readout of a single spin state in diamond via spin-to-charge
  conversion.
\newblock {\it Phys. Rev. Lett.\/} {\bf 114}, 136402 (2015).

\bibitem{Hausmann2011}
B.~J.~M. Hausmann, T.~M. Babinec, J.~T. Choy, J.~S. Hodges, S.~Hong, I.~Bulu,
  A.~Yacoby, M.~D. Lukin, M.~Lon{\v{c}}ar, Single-color centers implanted in
  diamond nanostructures.
\newblock {\it New J. Phys.\/} {\bf 13}, 045004 (2011).

\bibitem{BarGill2012}
N.~Bar-Gill, L.~M. Pham, C.~Belthangady, D.~Le~Sage, P.~Cappellaro, J.~R. Maze,
  M.~D. Lukin, A.~Yacoby, R.~Walsworth, Suppression of spin-bath dynamics for
  improved coherence of multi-spin-qubit systems.
\newblock {\it Nat. Commun.\/} {\bf 3}, 858 (2012).

\bibitem{Cohen2017}
I.~Cohen, N.~Aharon, A.~Retzker, Continuous dynamical decoupling utilizing
  time-dependent detuning.
\newblock {\it Fortschr. Phys.\/} {\bf 65}, 1600071 (2017).

\end{thebibliography}

\section*{Acknowledgments}
We thank R.-B. Liu for helpful discussions. \textbf{Funding:} This work was supported by the National Key Research and Development Program of China (Grants No. 2018YFA0306600 and 2016YFA0502400), the National Natural Science Foundation of China (Grants No. 81788101, 91636217, 11722544, 11761131011, and 31971156), the CAS (Grants No. GJJSTD20170001, QYZDY-SSW-SLH004 and YIPA2015370), the Anhui Initiative in Quantum Information Technologies (Grant No. AHY050000), the CEBioM, the national youth talent support program, the Fundamental Research Funds for the Central Universities, the China Postdoctoral Science Foundation (Grant No. BX20180294), and USTC Research Funds of the Double First-Class Initiative (Grant No. YD2340002004). \textbf{Author contributions:} J.D. and F.S. supervised the entire project. F.K., F.S. and J.D. conceived the experiments. P.Z. and F.K. prepared the setup and performed the experiments, together with Z.Q. and Z.W.. P.Y. and M.W. performed the ion implantation and nanofabrication. F.K. and Z.H. carried out the calculations. F.K., P.Z., and F.S. wrote the manuscript. All authors analysed the data, discussed the results and commented on the manuscript. \textbf{Competing interests:} The authors declare no competing interests. \textbf{Data and materials availability:} All data needed to evaluate the conclusions in the paper are present in the paper and/or the Supplementary Materials. Additional data related to this paper may be requested from the authors.

\end{document}


\section*{Supplementary Materials}

\subsection*{Section S1. Description of the sensor-target system}

The target spin consists of a spin-1/2 electron spin and a spin-1/2 nuclear spin. At zero magnetic field, only the hyperfine interaction is included, and thus the Hamiltonian can be written as
\begin{equation}
H_{\text{t}} = \mathbf{S}\cdot \mathbf{A} \cdot\mathbf{I},
\label{Htarget}
\end{equation}
where $\mathbf{A}$ is the hyperfine tensor, and is diagonal
\begin{equation}
\mathbf{A}=
\left(\begin{array}{ccc}
A_{\perp} &  & \\
 & A_{\perp} & \\
 &  & A_{\parallel}
\end{array}\right)
\end{equation}
in the principal axis frame. $\mathbf{S}$ and $\mathbf{I}$ are the electron and nuclear spin operators, respectively. To conveniently describe this system, we transform the bare spin-up and spin-down basis to the spin singlet and triplet basis $\{| T_{+1} \rangle, | S_{0} \rangle, | T_{0} \rangle, | T_{-1} \rangle\}$:
\begin{equation}
\begin{split}
&| T_{+1} \rangle = |\uparrow\uparrow\rangle, \\
&| S_{0} \rangle = \frac{1}{\sqrt{2}}(|\uparrow\downarrow\rangle-|\downarrow\uparrow\rangle), \\
&| T_{0} \rangle = \frac{1}{\sqrt{2}}(|\uparrow\downarrow\rangle+|\downarrow\uparrow\rangle), \\
&| T_{-1} \rangle = |\downarrow\downarrow\rangle, \\
\end{split}
\label{STstate}
\end{equation}
and then Eq.~\ref{Htarget} can be diagonalized as
\begin{equation}
H_{\text{t}}^{\mathcal{T}} = \mathcal{T}\cdot H_{\text{t}}\cdot\mathcal{T}^{-1}=
\frac{1}{4}\left(\begin{array}{cccc}
A_{\parallel} &  &  & \\
 & -A_{\parallel}-2A_{\perp} &  & \\
 &  & -A_{\parallel}+2A_{\perp}  & \\
 &  &  &  A_{\parallel}
\end{array}\right),
\label{HtargetR}
\end{equation}
where
\begin{equation}
\mathcal{T}=
\begin{pmatrix}
1 & 0 & 0 & 0 \\
0 & \frac{1}{\sqrt{2}} & -\frac{1}{\sqrt{2}} & 0 \\
0 & \frac{1}{\sqrt{2}} & \frac{1}{\sqrt{2}} & 0 \\
0 & 0 & 0 & 1
\end{pmatrix}
\label{R}
\end{equation}
is the transformation matrix. The eigenenergies of the target spin, $\omega_{T_{+1}}$, $\omega_{S_{0}}$, $\omega_{T_{0}}$, and $\omega_{T_{-1}}$, are just the diagonal elements in Eq.~\ref{HtargetR}. In this new basis, the electron spin operators can be written as
\begin{equation}
\begin{split}
& S_x^{\mathcal{T}} = \mathcal{T}\cdot(S_x \otimes I_2)\cdot\mathcal{T}^{-1}=
\frac{1}{2\sqrt{2}}\begin{pmatrix}
0 & -1 & 1 & 0\\
-1 & 0 & 0 & 1 \\
1 & 0 & 0 & 1 \\
0 & 1 & 1 & 0
\end{pmatrix},\\
& S_y^{\mathcal{T}} = \mathcal{T}\cdot(S_y \otimes I_2)\cdot\mathcal{T}^{-1}=
\frac{1}{2\sqrt{2}}\begin{pmatrix}
0 & \text{i} & -\text{i} & 0 \\
-\text{i} & 0 & 0 & -\text{i} \\
\text{i} & 0 & 0 &  -\text{i} \\
0 &  \text{i} & \text{i}  & 0
\end{pmatrix},\\
& S_z^{\mathcal{T}} = \mathcal{T}\cdot(S_z \otimes I_2)\cdot\mathcal{T}^{-1}=
\frac{1}{2}\begin{pmatrix}
1 & 0 & 0 & 0\\
0 & 0 & 1 & 0 \\
0 & 1 & 0 & 0\\
0 & 0 & 0 & -1
\end{pmatrix},
\end{split}
\label{SR}
\end{equation}
where $I_2$ is the $2\times2$ identity matrix. The nuclear spin operators can be transformed in a similar way, but are ignored due to the much smaller gyromagnetic ratio.

The Hamiltonian of magnetic noise can also be written in the new basis as
\begin{equation}
\delta H^{\mathcal{T}} = \sum_{j=x,y,z}\delta_j S_j^{\mathcal{T}}.
\label{noiseH}
\end{equation}
It can be diagonalized according to the perturbation theory:
\begin{equation}
\delta H_0^{\mathcal{T}} =
\begin{pmatrix}
\delta\omega_{T_{+1}} & 0 & 0 & 0\\
0 & \delta\omega_{S_{0}} & 0 & 0\\
0 & 0 & \delta\omega_{T_{0}} & 0\\
0 & 0 & 0 & \delta\omega_{T_{-1}}
\end{pmatrix},
\label{noiseH0}
\end{equation}
where the energy level shifts are
\begin{equation}
\begin{split}
& \delta\omega_{T_{+1}} = \frac{\delta_z}{2} + O(\delta^2), \\
& \delta\omega_{S_{0}} = -\frac{{\delta_x}^2+{\delta_y}^2}{2(A_{\parallel}+A_{\perp})}-\frac{{\delta_z}^2}{4A_{\perp}} + O(\delta^4), \\
& \delta\omega_{T_{0}} = -\frac{{\delta_x}^2+{\delta_y}^2}{2(A_{\parallel}-A_{\perp})}+\frac{{\delta_z}^2}{4A_{\perp}} + O(\delta^4), \\
& \delta\omega_{T_{-1}} = -\frac{\delta_z}{2} + O(\delta^2).
\end{split}
\label{energyshift}
\end{equation}

If we apply a radiofrequency (RF) pulse perpendicular to the principle axis with frequency $\omega_{\text{ST}_{\pm 1}}=|\omega_{S_{0}}-\omega_{T_{\pm 1}}|$, for example, of the form $H_{1,\text{t}} = 2\Omega_{x} \cos{(\omega_{\text{ST}_{\pm 1}}t)}S_x^{\mathcal{T}}$, then a transition between $| S_{0} \rangle \leftrightarrow | T_{\pm 1} \rangle$ will happen. The transition operator can be written as
\begin{equation}
U_{\text{ST}_{\pm 1}}^{\mathcal{T}}(\theta) =
\begin{pmatrix}
\frac{1+\cos{\frac{\theta}{2}}}{2} & \frac{\text{i}\sin{\frac{\theta}{2}}}{\sqrt{2}} & 0 & \frac{1-\cos{\frac{\theta}{2}}}{2}\\
\frac{\text{i}\sin{\frac{\theta}{2}}}{\sqrt{2}} & \cos{\frac{\theta}{2}} & 0 & -\frac{\text{i}\sin{\frac{\theta}{2}}}{\sqrt{2}}\\
0 & 0 & 1 & 0\\
\frac{1-\cos{\frac{\theta}{2}}}{2} & -\frac{\text{i}\sin{\frac{\theta}{2}}}{\sqrt{2}} & 0 & \frac{1+\cos{\frac{\theta}{2}}}{2}
\end{pmatrix},
\label{UxT}
\end{equation}
where $\theta = \Omega_{x}t_{\text{RF}}$, $t_{\text{RF}}$ is the pulse length. Specifically, the operators of $\pi$ and $2\pi$ pulses are
\begin{equation}
U_{\text{ST}_{\pm 1}}^{\mathcal{T}}(\pi) =
\begin{pmatrix}
\frac{1}{2} & \frac{\text{i}}{\sqrt{2}} & 0 & \frac{1}{2}\\
\frac{\text{i}}{\sqrt{2}} & 0 & 0 & -\frac{\text{i}}{\sqrt{2}}\\
0 & 0 & 1 & 0\\
\frac{1}{2} & -\frac{\text{i}}{\sqrt{2}} & 0 & \frac{1}{2}
\end{pmatrix},
U_{\text{ST}_{\pm 1}}^{\mathcal{T}}(2\pi) =
\begin{pmatrix}
0 & 0 & 0 & 1\\
0 & -1 & 0 & 0\\
0 & 0 & 1 & 0\\
1 & 0 & 0 & 0
\end{pmatrix}.
\label{UxPi}
\end{equation}
Similarly, a RF pulse parallel to the principle axis with frequency $\omega_{\text{ST}_{0}}=|\omega_{S_{0}}-\omega_{T_{0}}|$ will induce transition between $| S_{0} \rangle \leftrightarrow | T_{0} \rangle$, and the corresponding transition operator is
\begin{equation}
U_{\text{ST}_{0}}^{\mathcal{T}}(\theta) =
\begin{pmatrix}
1 & 0 & 0 & 0\\
0 & \cos{\frac{\theta}{2}} & -\text{i}\sin{\frac{\theta}{2}} & 0\\
0 & -\text{i}\sin{\frac{\theta}{2}} & \cos{\frac{\theta}{2}} & 0\\
0 & 0 & 0 & 1
\end{pmatrix}.
\label{UzT}
\end{equation}
The transition between $| T_{0} \rangle \leftrightarrow | T_{\pm 1} \rangle$ can also be driven by perpendicular RF with the corresponding resonant frequency, but has not been involved in our experiments.

At zero magnetic field, the Hamiltonian of NV centers can be written as
\begin{equation}
H_{\text{NV}} = D{S_{z}^{\text{NV}}}^{2},
\label{HNV}
\end{equation}
where $D=2.87$ GHz is the zero-field splitting of the NV center, $\mathbf{S}^{\text{NV}}$ is the spin-1 operator for the NV electron spin.

The dipole-dipole coupling between the NV center and the target spin can be written as
\begin{equation}
H_{\text{dd}} = \frac{\mu_{0}\gamma_{\text{NV}}\gamma_{\text{tar}}\hbar}{4\pi} [\frac{\mathbf{S}^{\text{NV}}\cdot\mathbf{S}^{\text{tar}}}{r^{3}}-
\frac{3(\mathbf{S}^{\text{NV}}\cdot\mathbf{r})(\mathbf{r}\cdot\mathbf{S}^{\text{tar}})}{r^{5}}],
\label{interaction}
\end{equation}
where $\gamma_{\text{NV}}$ and $\gamma_{\text{tar}}$ are the gyromagnetic ratios of the NV and target electron spin, respectively. $\mathbf{S}^{\text{NV}}$ and $\mathbf{S}^{\text{tar}}$ are the spin operators of the NV and target electron spin, respectively. $\mathbf{r}$ is the separation vector between the sensor and the target. Here all the vectors are defined in the NV frame with the $z$ axis towards the N-V axis, where $\mathbf{r}$ and the principal axis of the target spin are characterized by $\{\theta_{r},\phi_{r}\}$ and $\{\theta_{\text{e}},\phi_{\text{e}}\}$, respectively. The transformation from the principle axis frame to the NV frame can be represented by a rotation matrix:
\begin{equation}
\mathcal{R}=
\begin{pmatrix}
\cos{\theta_{\text{e}}}\cos{\phi_{\text{e}}} & -\sin{\phi_{\text{e}}} & \sin{\theta_{\text{e}}}\cos{\phi_{\text{e}}}\\
\cos{\theta_{\text{e}}}\sin{\phi_{\text{e}}} & \cos{\phi_{\text{e}}} & \sin{\theta_{\text{e}}}\sin{\phi_{\text{e}}}\\
-\sin{\theta_{\text{e}}} & 0 & \cos{\theta_{\text{e}}}
\end{pmatrix}.
\end{equation}
Then, we have
\begin{equation}
\mathbf{S}^{\text{tar}}=\mathcal{R}\cdot \mathbf{S}^{\mathcal{T}},
\label{frameR}
\end{equation}
here $\mathbf{S}^{\mathcal{T}}$ is defined in Eq.~\ref{SR}. Such coupling is usually a small perturbation to the Hamiltonian of the NV center (Eq.~\ref{HNV}) and the target spin (Eq.~\ref{HtargetR}), so the secular approximation can be performed, and Eq.~\ref{interaction} can be simplified to
\begin{equation}
\begin{split}
H_{\text{dd}} & \approx \frac{\mu_{0}\gamma_{\text{NV}}\gamma_{\text{tar}} \hbar}{4\pi r^3}(\cos{\theta_{\text{e}}}-3\cos{\theta_r}\cos{\theta_{r^{'}}})S_{z}^{\text{NV}}S_{zz}^{\mathcal{T}} \\
& = C(\theta_{\text{e}},r,\theta_r,\theta_{r^{'}})S_{z}^{\text{NV}}S_{zz}^{\mathcal{T}},
\end{split}
\label{simple dd}
\end{equation}
where $\theta_{r^{'}}$ is the angle between $\mathbf{r}$ and the principal axis, and $S_{zz}^{\mathcal{T}}$ is a simplified form of $S_{z}^{\mathcal{T}}$:
\begin{equation}
S_{zz}^{\mathcal{T}} =
\frac{1}{2}\begin{pmatrix}
1 &   &   &  \\
  & 0 &   &   \\
  &   & 0 &  \\
  &   &   & -1
\end{pmatrix}
\label{SzzR}
\end{equation}

\newpage

\subsection*{Section S2. Calculations of DEER signal}

The DEER signal is determined by the accumulated phase $\phi = a\cdot \tau$, where $a$ is the dipolar coupling strength, depending on the state of the target spin according to Eq.~\ref{simple dd}. The target spin is in thermal equilibrium, which means it has nearly equal probability in $| T_{+1} \rangle$, $| S_{0} \rangle$, $| T_{0} \rangle$, and $| T_{-1} \rangle$ states. To quantitatively describe the zero-field DEER signal, we disassemble the evolution as following:
\begin{table}[H]
\centering
\renewcommand\arraystretch{2.0}
  \begin{tabular}{ c  c | c }
    \hline
    \multicolumn{2}{c|}{Transition} & Accumulated phase $\phi$ \\
    \hline
    \multirow{3}*{$| T_{+1} \rangle \xrightarrow{U_{\text{ST}_{\pm 1}}^{\mathcal{T}}(\theta)}$} & $(\frac{1}{2}+\frac{1}{2}\cos{\frac{\theta}{2}})| T_{+1} \rangle$ & 0 \\
    ~ & $+\frac{\text{i}}{\sqrt{2}}\sin{\frac{\theta}{2}}| S_{0} \rangle$ & $\frac{1}{4}C\tau$ \\
    ~ & $+(\frac{1}{2}-\frac{1}{2}\cos{\frac{\theta}{2}})| T_{-1} \rangle$ & $\frac{1}{2}C\tau$ \\
    \hline
    \multirow{3}*{$| S_{0} \rangle \xrightarrow{U_{\text{ST}_{\pm 1}}^{\mathcal{T}}(\theta)}$} & $\frac{\text{i}}{\sqrt{2}}\sin{\frac{\theta}{2}}| T_{+1} \rangle$ & $-\frac{1}{4}C\tau$ \\
    ~ & $+\cos{\frac{\theta}{2}}| S_{0} \rangle$ & 0 \\
    ~ & $-\frac{\text{i}}{\sqrt{2}}\sin{\frac{\theta}{2}}| T_{-1} \rangle$ & $\frac{1}{4}C\tau$ \\
    \hline
    \multirow{1}*{$| T_{0} \rangle \xrightarrow{U_{\text{ST}_{\pm 1}}^{\mathcal{T}}(\theta)}$} & $| T_{0} \rangle$ & 0 \\
    \hline
    \multirow{3}*{$| T_{-1} \rangle \xrightarrow{U_{\text{ST}_{\pm 1}}^{\mathcal{T}}(\theta)}$} & $(\frac{1}{2}-\frac{1}{2}\cos{\frac{\theta}{2}})| T_{+1} \rangle$ & $-\frac{1}{2}C\tau$ \\
    ~ & $+\frac{\text{i}}{\sqrt{2}}\sin{\frac{\theta}{2}}| S_{0} \rangle$ & $-\frac{1}{4}C\tau$ \\
    ~ & $+(\frac{1}{2}+\frac{1}{2}\cos{\frac{\theta}{2}})| T_{-1} \rangle$ & 0 \\
    \hline
  \end{tabular}
\end{table}
The signal can be calculated as the weighted average of each rows, which is
\small
\begin{equation}
\begin{split}
S(\theta,\tau) & = \langle \cos^{2}{\phi} \rangle \\
& = \frac{1}{32}[25+3\cos{\theta}+4\cos{\frac{\theta}{2}}+4(1-\cos{\theta})\cos{\frac{C\tau}{2}}+2(1-\cos{\frac{\theta}{2}})^2 \cos{C\tau}] \\
& = \frac{1}{32}[25+3\cos{C\tau}+4\cos{\frac{C\tau}{2}}+4(1-\cos{C\tau})\cos{\frac{\theta}{2}}+2(1-\cos{\frac{C\tau}{2}})^2 \cos{\theta}].
\end{split}
\label{DEER_signal}
\end{equation}
\normalsize
In general, one can see an dual-frequency oscillation by varying the RF pulse length, i.e., $\theta$, with fixed evolution time $\tau$, and vice versa. However, a better strategy is utilizing the faster term in $\tau$ domain because of the decoherence process during the evolution. So we choose $2\pi$ RF pulse rather than $\pi$ RF pulse, and then the zero-field DEER signal can be simplified as $(3+\cos{C\tau})/4$. Due to the decoherence, there will be an extra random phase $\delta \phi(\tau)$ accumulated during the evolution. The average effect of this random phase is a stretched exponential decay, and thus the signal can be written as
\begin{equation}
S(\tau) = \frac{3}{4}+\frac{1}{4}e^{-(\tau/T_{2,\text{NV}})^p}\cos{C\tau},
\label{DEER_signal2}
\end{equation}
where $p$ is in the range of $1-3$ determined by the dynamic of bath.

\newpage

\subsection*{Section S3. Correlation Rabi measurement}

To see how the manipulation of the target spin during spin-locking period (the black dash box in Fig.~2a in the main text) affects the correlation signal, we use a similar strategy described above to disassemble the evolution. First, we write the operator of the manipulation. For example, the operator corresponding to $\text{ST}_{\pm1}$ Rabi oscillation (Fig.~2b in the main text) is $U_{\text{Rabi},\text{ST}_{\pm 1}} = U_{\text{ST}_{\pm 1}}^{\mathcal{T}}(\theta)$ given by Eq.~\ref{UxT}. Then, the evolution is disassembled:
\begin{table}[H]
\centering
\renewcommand\arraystretch{2.0}
  \begin{tabular}{ c c l | c | c}
    \hline
    \multicolumn{3}{c|}{Evolution of the target spin} & $\phi_1$ & $\phi_2$\\
    \hline
    \multirow{3}*{$| T_{+1} \rangle \xrightarrow{2\pi} | T_{-1} \rangle \xrightarrow{U}$} & $(\frac{1}{2}-\frac{1}{2}\cos{\frac{\theta}{2}})| T_{+1} \rangle$ & $\xrightarrow{2\pi} | T_{-1} \rangle$ & $\frac{1}{2}C\tau$ & $\frac{1}{2}C\tau$ \\
    ~ & $+\frac{\text{i}}{\sqrt{2}}\sin{\frac{\theta}{2}}| S_{0} \rangle$ & $\xrightarrow{2\pi} | S_{0} \rangle$ & $\frac{1}{2}C\tau$ & 0 \\
    ~ & $+(\frac{1}{2}+\frac{1}{2}\cos{\frac{\theta}{2}})| T_{-1} \rangle$ & $\xrightarrow{2\pi} | T_{+1} \rangle$ & $\frac{1}{2}C\tau$ & $-\frac{1}{2}C\tau$ \\
    \hline
    \multirow{3}*{$| S_{0} \rangle \xrightarrow{2\pi} | S_{0} \rangle \xrightarrow{U}$} & $+\frac{\text{i}}{\sqrt{2}}\sin{\frac{\theta}{2}}| T_{+1} \rangle$ & $\xrightarrow{2\pi} | T_{-1} \rangle$ & 0 & $\frac{1}{2}C\tau$ \\
    ~ & $+\cos{\frac{\theta}{2}}| S_{0} \rangle$ & $\xrightarrow{2\pi} | S_{0} \rangle$ & 0 & 0 \\
    ~ & $-\frac{\text{i}}{\sqrt{2}}\sin{\frac{\theta}{2}}| T_{-1} \rangle$ & $\xrightarrow{2\pi} | T_{+1} \rangle$ & 0 & $-\frac{1}{2}C\tau$ \\
    \hline
    \multirow{1}*{$| T_{0} \rangle \xrightarrow{2\pi} | T_{0} \rangle \xrightarrow{U}$} & $| T_{0} \rangle$ & $\xrightarrow{2\pi} | T_{0} \rangle$ & 0 & 0 \\
    \hline
    \multirow{3}*{$| T_{-1} \rangle \xrightarrow{2\pi} | T_{+1} \rangle \xrightarrow{U}$} & $(\frac{1}{2}+\frac{1}{2}\cos{\frac{\theta}{2}})| T_{+1} \rangle$ & $\xrightarrow{2\pi} | T_{-1} \rangle$ & $-\frac{1}{2}C\tau$ & $\frac{1}{2}C\tau$ \\
    ~ & $+\frac{\text{i}}{\sqrt{2}}\sin{\frac{\theta}{2}}| S_{0} \rangle$ & $\xrightarrow{2\pi} | S_{0} \rangle$ & $-\frac{1}{2}C\tau$ & 0 \\
    ~ & $+(\frac{1}{2}-\frac{1}{2}\cos{\frac{\theta}{2}})| T_{-1} \rangle$ & $\xrightarrow{2\pi} | T_{+1} \rangle$ & $-\frac{1}{2}C\tau$ & $-\frac{1}{2}C\tau$ \\
    \hline
  \end{tabular}
\end{table}
The corresponding correlation signal is
\small
\begin{equation}
\boxed{
S_{\text{Rabi},\text{ST}_{\pm 1}} \sim \langle \cos{2\phi_1}\cos{2\phi_2} \rangle = \frac{1}{8}(1-\cos{C\tau})^2\cos{\theta} + \frac{3}{8} + \frac{1}{4}\cos{C\tau} + \frac{3}{8}\cos^2{C\tau},
}
\end{equation}
\normalsize
which is a single-frequency oscillation. The operator corresponding to $\text{ST}_{0}$ Rabi oscillation (Fig.~2c in the main text) is
\small
\begin{equation}
U_{\text{Rabi},\text{ST}_{0}} = U_{\text{ST}_{\pm 1}}^{\mathcal{T}}(\pi) \cdot U_{\text{ST}_{0}}^{\mathcal{T}}(\theta) \cdot U_{\text{ST}_{\pm 1}}^{\mathcal{T}}(\pi)=
\begin{pmatrix}
\frac{1-\cos{\frac{\theta}{2}}}{2} & 0 & \frac{\sin{\frac{\theta}{2}}}{\sqrt{2}} & \frac{1+\cos{\frac{\theta}{2}}}{2}\\
0 & -1 & 0 & 0\\
\frac{\sin{\frac{\theta}{2}}}{\sqrt{2}} & 0 & \cos{\frac{\theta}{2}} & -\frac{\sin{\frac{\theta}{2}}}{\sqrt{2}}\\
\frac{1+\cos{\frac{\theta}{2}}}{2} & 0 & -\frac{\sin{\frac{\theta}{2}}}{\sqrt{2}} & \frac{1-\cos{\frac{\theta}{2}}}{2}
\end{pmatrix},
\end{equation}
\normalsize
and the correlation signal can be calculated in a similar way, which is
\small
\begin{equation}
\boxed{
S_{\text{Rabi},\text{ST}_{0}} \sim \langle \cos{2\phi_1}\cos{2\phi_2} \rangle = \frac{1}{8}(1-\cos{C\tau})^2\cos{\theta} + \frac{3}{8} + \frac{1}{4}\cos{C\tau} + \frac{3}{8}\cos^2{C\tau}.
}
\end{equation}
\normalsize

\newpage

\subsection*{Section S4. Correlation Ramsey measurement}

The operators corresponding to $\text{ST}_{\pm 1}$ Ramsey measurement (Fig.~3a in the main text) are
\small
\begin{equation}
\begin{split}
U_{\text{Ramsey},\text{ST}_{\pm 1}}^{\text{sig}} & = U_{\text{ST}_{\pm 1}}^{\mathcal{T}}(-\frac{\pi}{2}) \cdot \exp[-\text{i}(H_{\text{t}}^{\mathcal{T}}+\delta H_0^{\mathcal{T}}) t] \cdot U_{\text{ST}_{\pm 1}}^{\mathcal{T}}(\frac{\pi}{2}) \\
& = U_{\text{phase}}\cdot\begin{pmatrix}
\frac{1}{4}(1+\alpha e^{-\text{i}\varphi}) & \frac{\text{i}\sqrt{2}}{4}(-1+\beta e^{-\text{i}\varphi}) & 0 & \frac{1}{4}(-1+\gamma e^{-\text{i}\varphi}) \\
\frac{\text{i}\sqrt{2}}{4}(-1+\beta e^{-\text{i}\varphi}) & \frac{1}{2}(1+\gamma e^{-\text{i}\varphi}) & 0 & \frac{\text{i}\sqrt{2}}{4}(-1+\beta^* e^{-\text{i}\varphi})\\
0 & 0 & 1 & 0\\
\frac{1}{4}(-1+\gamma e^{-\text{i}\varphi}) & \frac{\text{i}\sqrt{2}}{4}(-1+\beta^* e^{-\text{i}\varphi}) & 0 & \frac{1}{4}(1+\alpha^* e^{-\text{i}\varphi})
\end{pmatrix},
\end{split}
\end{equation}
\normalsize
and
\small
\begin{equation}
\begin{split}
U_{\text{Ramsey},\text{ST}_{\pm 1}}^{\text{ref}} & = U_{\text{ST}_{\pm 1}}^{\mathcal{T}}(\frac{\pi}{2}) \cdot \exp[-\text{i}(H_{\text{t}}^{\mathcal{T}}+\delta H_0^{\mathcal{T}}) t] \cdot U_{\text{ST}_{\pm 1}}^{\mathcal{T}}(\frac{\pi}{2}) \\
& = U_{\text{phase}}\cdot\begin{pmatrix}
\frac{1}{4}(-1+\alpha e^{-\text{i}\varphi}) & \frac{\text{i}\sqrt{2}}{4}(1+\beta e^{-\text{i}\varphi}) & 0 & \frac{1}{4}(1+\gamma e^{-\text{i}\varphi}) \\
\frac{\text{i}\sqrt{2}}{4}(1+\beta e^{-\text{i}\varphi}) & \frac{1}{2}(1-\gamma e^{-\text{i}\varphi}) & 0 & \frac{\text{i}\sqrt{2}}{4}(-1-\beta^* e^{-\text{i}\varphi})\\
0 & 0 & 1 & 0\\
\frac{1}{4}(1+\gamma e^{-\text{i}\varphi}) & \frac{\text{i}\sqrt{2}}{4}(-1-\beta^* e^{-\text{i}\varphi}) & 0 & \frac{1}{4}(-1+\alpha^* e^{-\text{i}\varphi})
\end{pmatrix},
\end{split}
\end{equation}
\normalsize
where $H_{\text{t}}^{\mathcal{T}}$ and $\delta H_0^{\mathcal{T}}$ are given by Eq.~\ref{HtargetR} and Eq.~\ref{noiseH0}, respectively, the phase operator $U_{\text{phase}}$ can be ignored, $\varphi = \omega_{\text{ST}_{\pm 1}}t$, and the three coefficients are
\small
\begin{equation}
\begin{split}
\alpha & = 3\cos(\delta \omega_{\text{ST}_{\pm 1}} t)-\text{i}2\sqrt{2}\sin(\delta \omega_{\text{ST}_{\pm 1}} t), \\
\beta & = \cos(\delta \omega_{\text{ST}_{\pm 1}} t)-\text{i}\sqrt{2}\sin(\delta \omega_{\text{ST}_{\pm 1}} t), \\
\gamma & = \cos(\delta \omega_{\text{ST}_{\pm 1}} t).
\end{split}
\end{equation}
\normalsize
Then, the signal and reference are
\small
\begin{equation}
\begin{split}
S_{\text{Ramsey},\text{ST}_{\pm 1}}^{\text{sig}} & \sim \langle \cos{2\phi_1}\cos{2\phi_2} \rangle \\
& = \frac{1}{8}(1-\cos{C\tau})^2 \langle \cos(\delta \omega_{\text{ST}_{\pm 1}} t)\rangle \cos(\omega_{\text{ST}_{\pm 1}}t) \\
& + \frac{1}{32}(1-\cos{C\tau})^2 \langle\cos(2\delta \omega_{\text{ST}_{\pm 1}} t)\rangle + \frac{11}{32} + \frac{5}{16}\cos{C\tau} + \frac{11}{32}\cos^2{C\tau},
\end{split}
\end{equation}
\normalsize
and
\small
\begin{equation}
\begin{split}
S_{\text{Ramsey},\text{ST}_{\pm 1}}^{\text{ref}} & \sim \langle \cos{2\phi_1}\cos{2\phi_2} \rangle \\
& = -\frac{1}{8}(1-\cos{C\tau})^2 \langle \cos(\delta \omega_{\text{ST}_{\pm 1}} t)\rangle \cos(\omega_{\text{ST}_{\pm 1}}t) \\
& + \frac{1}{32}(1-\cos{C\tau})^2 \langle\cos(2\delta \omega_{\text{ST}_{\pm 1}} t)\rangle + \frac{11}{32} + \frac{5}{16}\cos{C\tau} + \frac{11}{32}\cos^2{C\tau},
\end{split}
\end{equation}
\normalsize
respectively. The measured differential signal is
\small
\begin{equation}
\boxed{
S_{\text{Ramsey},\text{ST}_{\pm 1}} = S_{\text{Ramsey},\text{ST}_{\pm 1}}^{\text{sig}}-S_{\text{Ramsey},\text{ST}_{\pm 1}}^{\text{ref}} \sim  \frac{1}{4}(1-\cos{C\tau})^2 \langle \cos(\delta \omega_{\text{ST}_{\pm 1}} t)\rangle \cos(\omega_{\text{ST}_{\pm 1}}t).
}
\end{equation}
\normalsize
Similarly, the operators corresponding to $\text{ST}_{0}$ Ramsey measurement (Fig.~3b in the main text) is
\small
\begin{equation}
\begin{split}
U_{\text{Ramsey},\text{ST}_{0}}^{\text{sig}} & =U_{\text{ST}_{\pm 1}}^{\mathcal{T}}(\pi) \cdot U_{\text{ST}_{0}}^{\mathcal{T}}(-\frac{\pi}{2}) \cdot \exp[-\text{i}(H_{\text{t}}^{\mathcal{T}}+\delta H_0^{\mathcal{T}}) t] \cdot U_{\text{ST}_{0}}^{\mathcal{T}}(\frac{\pi}{2}) \cdot U_{\text{ST}_{\pm 1}}^{\mathcal{T}}(\pi) \\
& = U_{\text{phase}}\cdot\begin{pmatrix}
\frac{1}{2}(\xi \cos{\delta \varphi}-\cos{\frac{\varphi{'}}{2}}) & \frac{1}{\sqrt{2}}\xi \sin{\delta \varphi} & \frac{\text{i}}{\sqrt{2}} \sin{\frac{\varphi{'}}{2}} & \frac{1}{2}(\xi \cos{\delta \varphi}+\cos{\frac{\varphi{'}}{2}}) \\
\frac{1}{\sqrt{2}}\xi \sin{\delta \varphi} & -\xi \cos{\delta \varphi} & 0 & \frac{1}{\sqrt{2}}\xi \sin{\delta \varphi}\\
-\frac{\text{i}}{\sqrt{2}} \sin{\frac{\varphi{'}}{2}} & 0 & \cos{\frac{\varphi{'}}{2}} & \frac{\text{i}}{\sqrt{2}} \sin{\frac{\varphi{'}}{2}}\\
\frac{1}{2}(\xi \cos{\delta \varphi}+\cos{\frac{\varphi{'}}{2}}) & \frac{1}{\sqrt{2}}\xi \sin{\delta \varphi} & -\frac{\text{i}}{\sqrt{2}} \sin{\frac{\varphi{'}}{2}} & \frac{1}{2}(\xi \cos{\delta \varphi}-\cos{\frac{\varphi{'}}{2}})
\end{pmatrix},
\end{split}
\end{equation}
\normalsize
and
\small
\begin{equation}
\begin{split}
U_{\text{Ramsey},\text{ST}_{0}}^{\text{ref}} & =U_{\text{ST}_{\pm 1}}^{\mathcal{T}}(\pi) \cdot U_{\text{ST}_{0}}^{\mathcal{T}}(\frac{\pi}{2}) \cdot \exp[-\text{i}(H_{\text{t}}^{\mathcal{T}}+\delta H_0^{\mathcal{T}}) t] \cdot U_{\text{ST}_{0}}^{\mathcal{T}}(\frac{\pi}{2}) \cdot U_{\text{ST}_{\pm 1}}^{\mathcal{T}}(\pi) \\
& = U_{\text{phase}}\cdot\begin{pmatrix}
\frac{1}{2}(\zeta \cos{\delta \varphi}-\eta\sin{\frac{\varphi{'}}{2}}) & \frac{1}{\sqrt{2}}\xi \sin{\delta \varphi} & \frac{1}{\sqrt{2}} \cos{\frac{\varphi{'}}{2}} & \frac{1}{2}(\zeta \cos{\delta \varphi}+\eta\sin{\frac{\varphi{'}}{2}}) \\
\frac{1}{\sqrt{2}}\xi \sin{\delta \varphi} & -\xi \cos{\delta \varphi} & 0 & \frac{1}{\sqrt{2}}\xi \sin{\delta \varphi}\\
\frac{1}{\sqrt{2}} \cos{\frac{\varphi{'}}{2}} & 0 & -\text{i}\sin{\frac{\varphi{'}}{2}} & -\frac{1}{\sqrt{2}} \cos{\frac{\varphi{'}}{2}}\\
\frac{1}{2}(\zeta \cos{\delta \varphi}+\eta\sin{\frac{\varphi{'}}{2}}) & \frac{1}{\sqrt{2}}\xi \sin{\delta \varphi} & \frac{1}{\sqrt{2}} \cos{\frac{\varphi{'}}{2}} & \frac{1}{2}(\zeta \cos{\delta \varphi}-\eta\sin{\frac{\varphi{'}}{2}})
\end{pmatrix},
\end{split}
\end{equation}
\normalsize
where $\varphi{'} = (\omega_{\text{ST}_{0}}+\delta \omega_{\text{ST}_{0}})t$, $\delta \varphi = 2\delta \omega_{\text{ST}_{\pm 1}} t$, $\xi$, $\zeta$, and $\eta$ are phase factors, which can be ignored. Then, the signal and reference are
\small
\begin{equation}
\begin{split}
S_{\text{Ramsey},\text{ST}_{0}}^{\text{sig}} & \sim \langle \cos{2\phi_1}\cos{2\phi_2} \rangle \\
& = \frac{1}{8}(1-\cos{C\tau})^2 \langle \cos(\omega_{\text{ST}_{0}} + \delta \omega_{\text{ST}_{0}} t)\rangle \\
& + \frac{1}{8}(1-\cos{C\tau})^2 \langle\cos(2\delta \omega_{\text{ST}_{\pm 1}} t)\rangle + \frac{1}{4}(1+\cos{C\tau})^2,
\end{split}
\end{equation}
\normalsize
and
\small
\begin{equation}
\begin{split}
S_{\text{Ramsey},\text{ST}_{0}}^{\text{ref}} & \sim \langle \cos{2\phi_1}\cos{2\phi_2} \rangle \\
& = -\frac{1}{8}(1-\cos{C\tau})^2 \langle \cos(\omega_{\text{ST}_{0}} + \delta \omega_{\text{ST}_{0}} t)\rangle \\
& + \frac{1}{8}(1-\cos{C\tau})^2 \langle\cos(2\delta \omega_{\text{ST}_{\pm 1}} t)\rangle + \frac{1}{4}(1+\cos{C\tau})^2,
\end{split}
\end{equation}
\normalsize
respectively. The measured differential signal is
\small
\begin{equation}
\boxed{
S_{\text{Ramsey},\text{ST}_{0}} = S_{\text{Ramsey},\text{ST}_{0}}^{\text{sig}}-S_{\text{Ramsey},\text{ST}_{0}}^{\text{ref}} \sim  \frac{1}{4}(1-\cos{C\tau})^2 \langle \cos(\omega_{\text{ST}_{0}} + \delta \omega_{\text{ST}_{0}} t)\rangle.
}
\end{equation}
\normalsize

\newpage

\begin{figure}
\centering
\includegraphics[width=1\columnwidth]{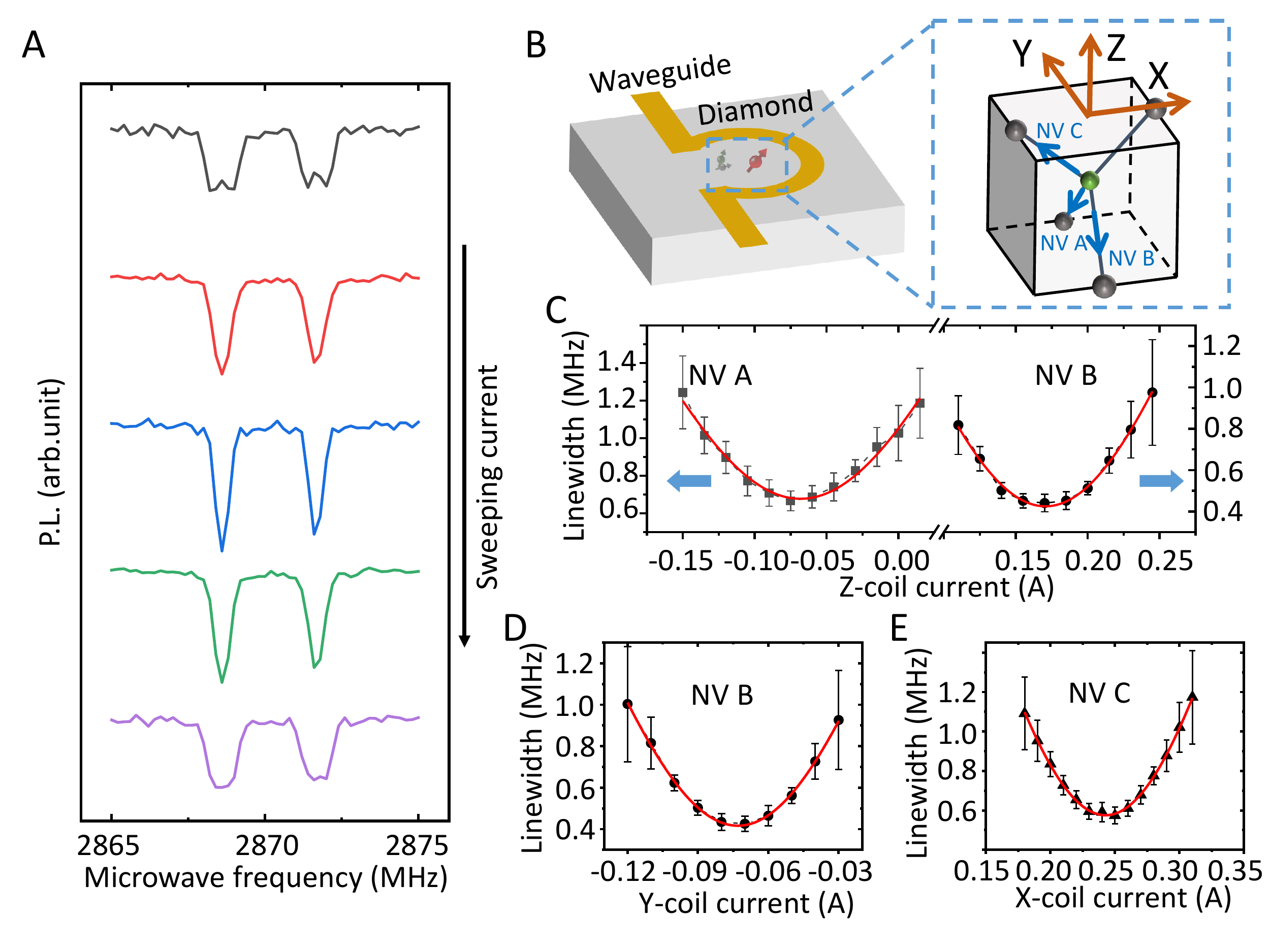} \caption{\textbf{Schematics of the compensation process.}
(\textbf{A}) Variations of the ODMR spectra with sweeping currents applied to Helmholtz coils.
(\textbf{B}) Orientations of four kinds of N-C bonds with respect to the lab frame. The blue arrows indicate the three differently orientated NV centers used for the magnetic-field compensation.
(\textbf{C})(\textbf{D})(\textbf{E}) Compensation of $B_z$, $B_y$ and $B_x$, respectively. Series of ODMR spectra are measured with sweeping currents, and the linewidths are extracted with Gaussian fitting (not shown). The points are the fitting linewithds with error bars indicating the fitting error. To find the symmetric center points, a symmetric-peak function is used to fit the data. Here we choose Gaussian function for simplicity. The solid lines are fitting results.
}
\label{Zerofield}
\end{figure}

\begin{figure}
\centering
\includegraphics[width=1\columnwidth]{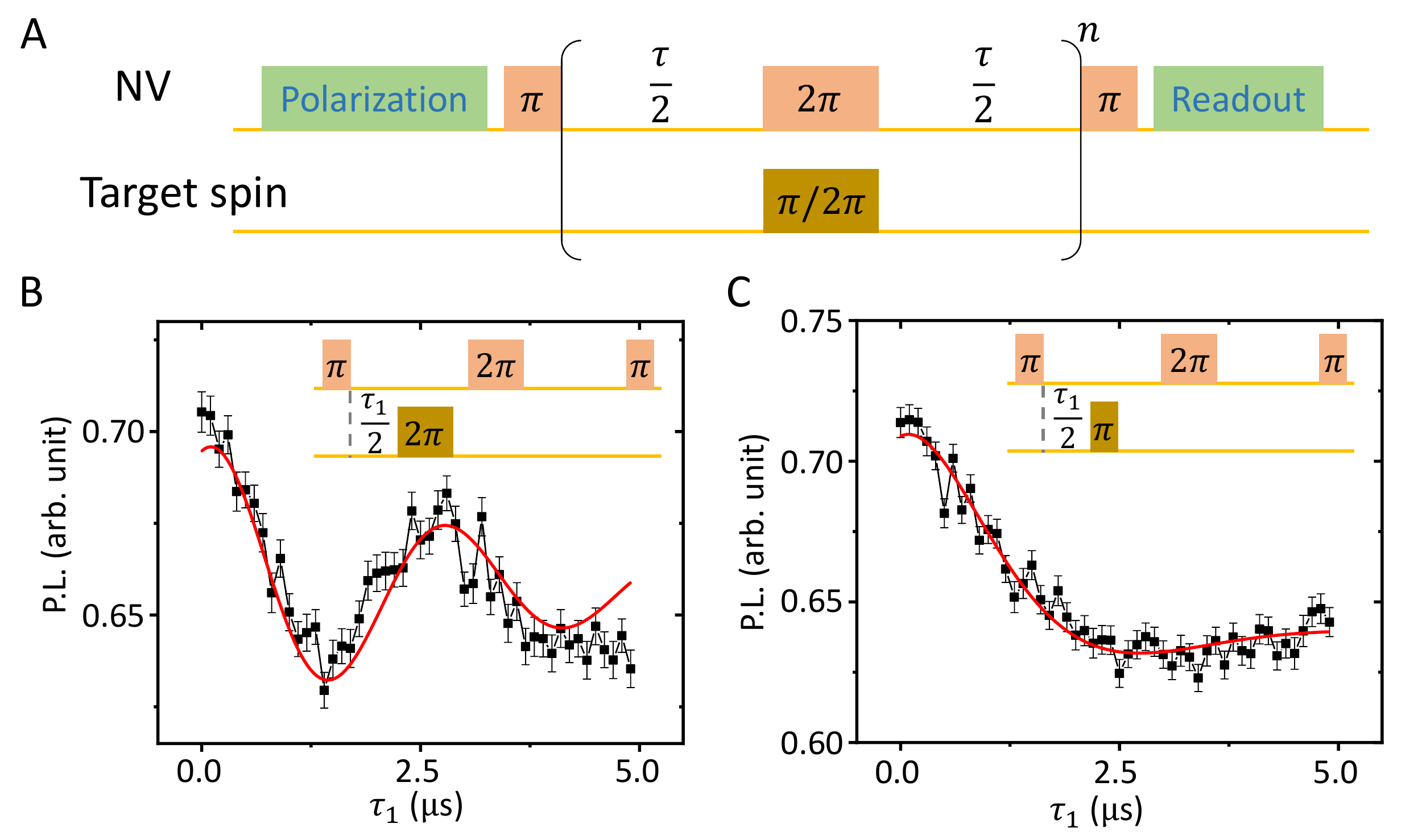} \caption{\textbf{Zero-field DEER measurement.}
(\textbf{A}) The zero-field DEER sequence. The sequence is similar to the ordinary non-zero field case, but the $\pi/2$ and $\pi$ pulses are replaced by $\pi$ and $2\pi$ pulses, respectively.
(\textbf{B})(\textbf{C}) Comparison between two DEER measurements. The difference is flipping the target spin with $2\pi$ (A) or $\pi$ (B) RF pulse. Here $\tau=8$ $\mu$s. A faster evolution is observed by applying $2\pi$ RF pulse. The points are experimental results, while the lines are sine decay fitting. Error bars indicate s.e.m. Note here the `decay' is actually a modulation, which is induced by multiple coupling strengths between the NV center and the P1 center, because of the fast jump of the P1 orientations.
}
\label{DEERseq}
\end{figure}

\begin{figure}
\centering
\includegraphics[width=1\columnwidth]{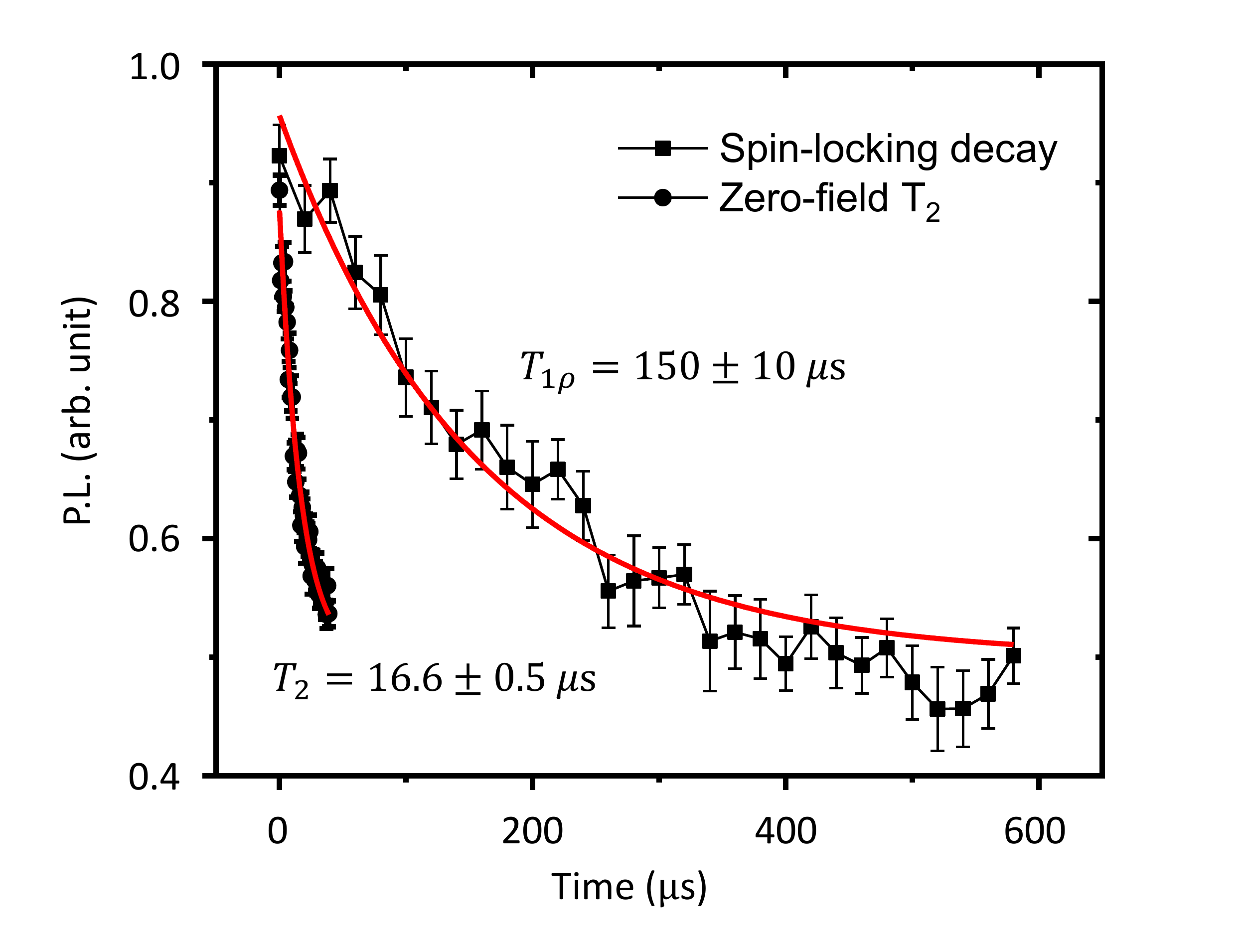} \caption{\textbf{Coherence properties of the NV center.}
The $T_2$ is measured by a similar sequence as the zero-field DEER, but without the manipulations on the target spin. The spin-locking decay is measured by applying a continuous driving field of the form $\Omega_1 \cos{[Dt+2\Omega_2/\Omega_1\sin{\Omega_1 t}]}$, where $D = 2.87$ GHz, $\Omega_1 = 30$ MHz, and $\Omega_2 = 9$ MHz. In general, an oscillation with frequency of $\Omega_1$ can be observed according the calculations in Materials and Methods. Here we sample the data at integral periods, and thus only the envelope is presented. The points and lines are experimental and exponential fitting results, respectively. Error bars indicate s.e.m.
}
\label{Coherence}
\end{figure}

\begin{figure}
\centering
\includegraphics[width=1\columnwidth]{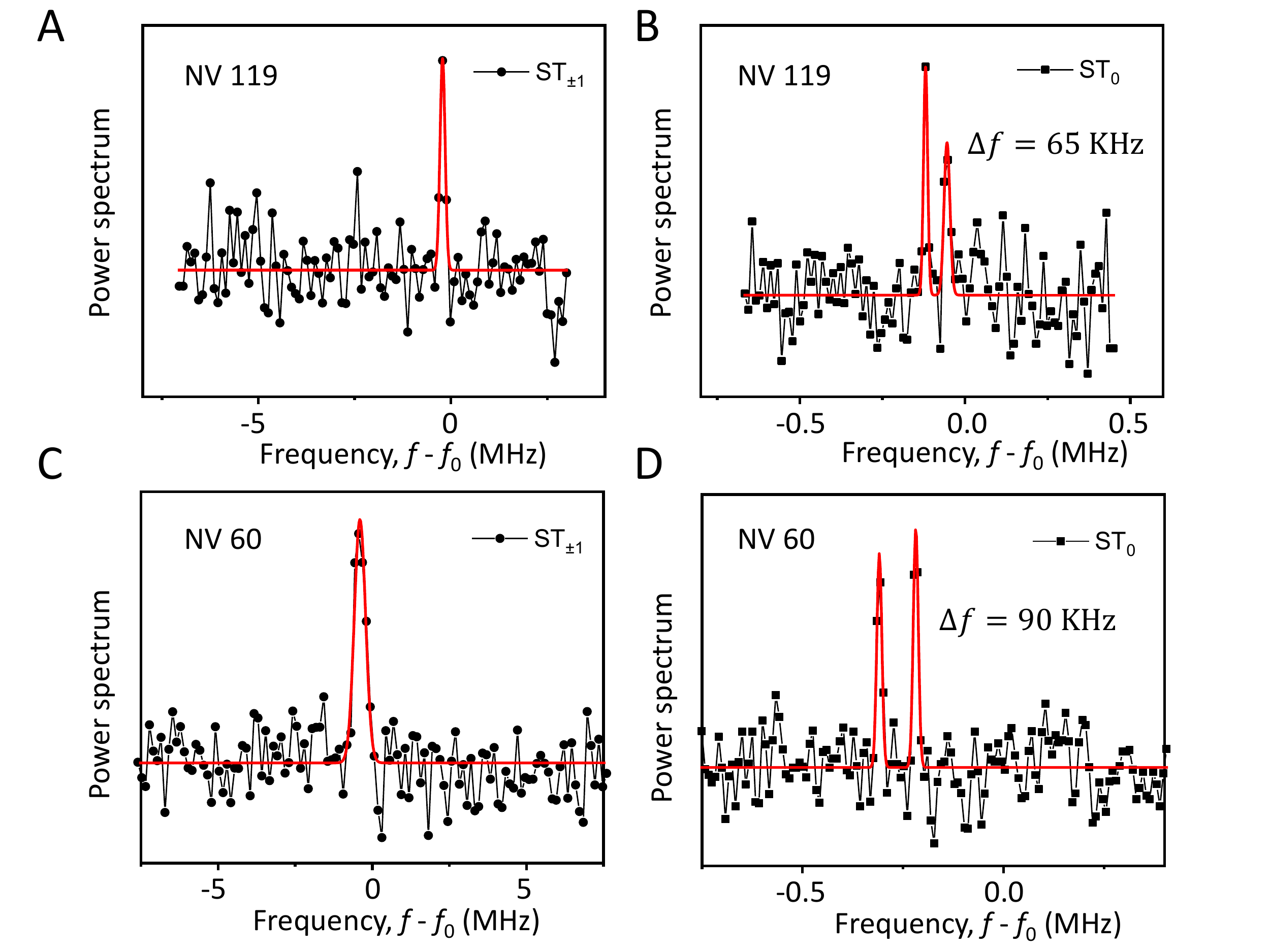} \caption{\textbf{Zero-field EPR spectra of different P1 centers.}
(\textbf{A})(\textbf{C}) $\text{ST}_{\pm 1}$ spectra. No obvious line splitting can be seen.
(\textbf{B})(\textbf{D}) $\text{ST}_{0}$ spectra. Clear line splitting can be seen with different splitting values.
The data are measured on a $^{12}$C isotropically purified diamond, leading to the the disappearance of the line splitting in $\text{ST}_{\pm 1}$ spectra. However, the line splitting in $\text{ST}_{0}$ spectra remains, which rule out the reason of magnetic dipolar coupling with nearby nuclear spins. This splitting also differs for different P1 centers, suggesting different local electric or strain environments.
}
\label{spectra}
\end{figure}

\begin{figure}
\centering
\includegraphics[width=1\columnwidth]{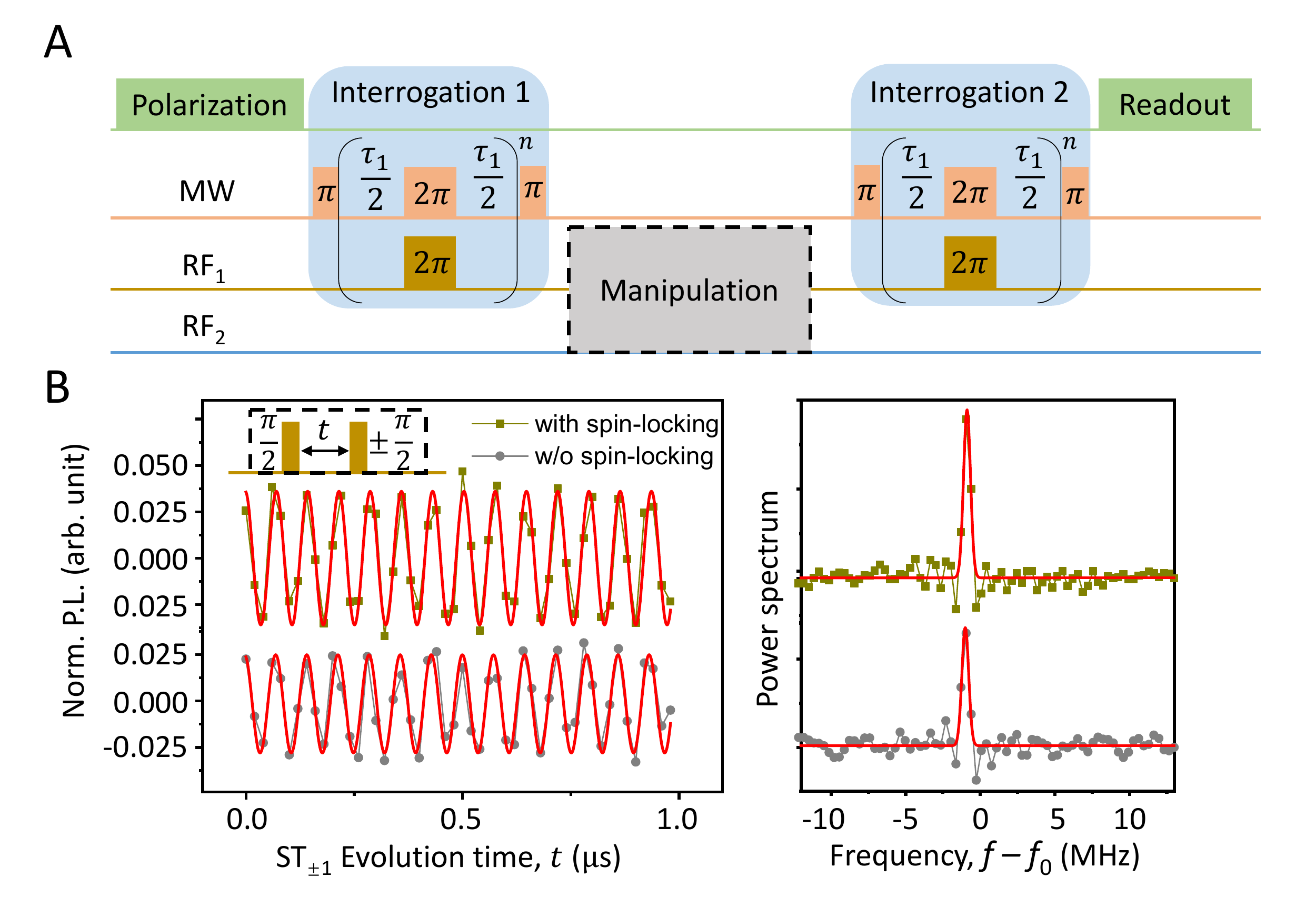} \caption{\textbf{An alternative protocol for correlation detection.}
(\textbf{A}) The pulse sequence. Spin-locking sequence is replaced by two $\pi$ pulses.
(\textbf{B}) Comparison between the two correlation detection protocol. The signal contrast of Ramsey experiment is partially lost without spin locking.
}
\label{T1Correlation}
\end{figure}

\begin{figure}
\centering
\includegraphics[width=1\columnwidth]{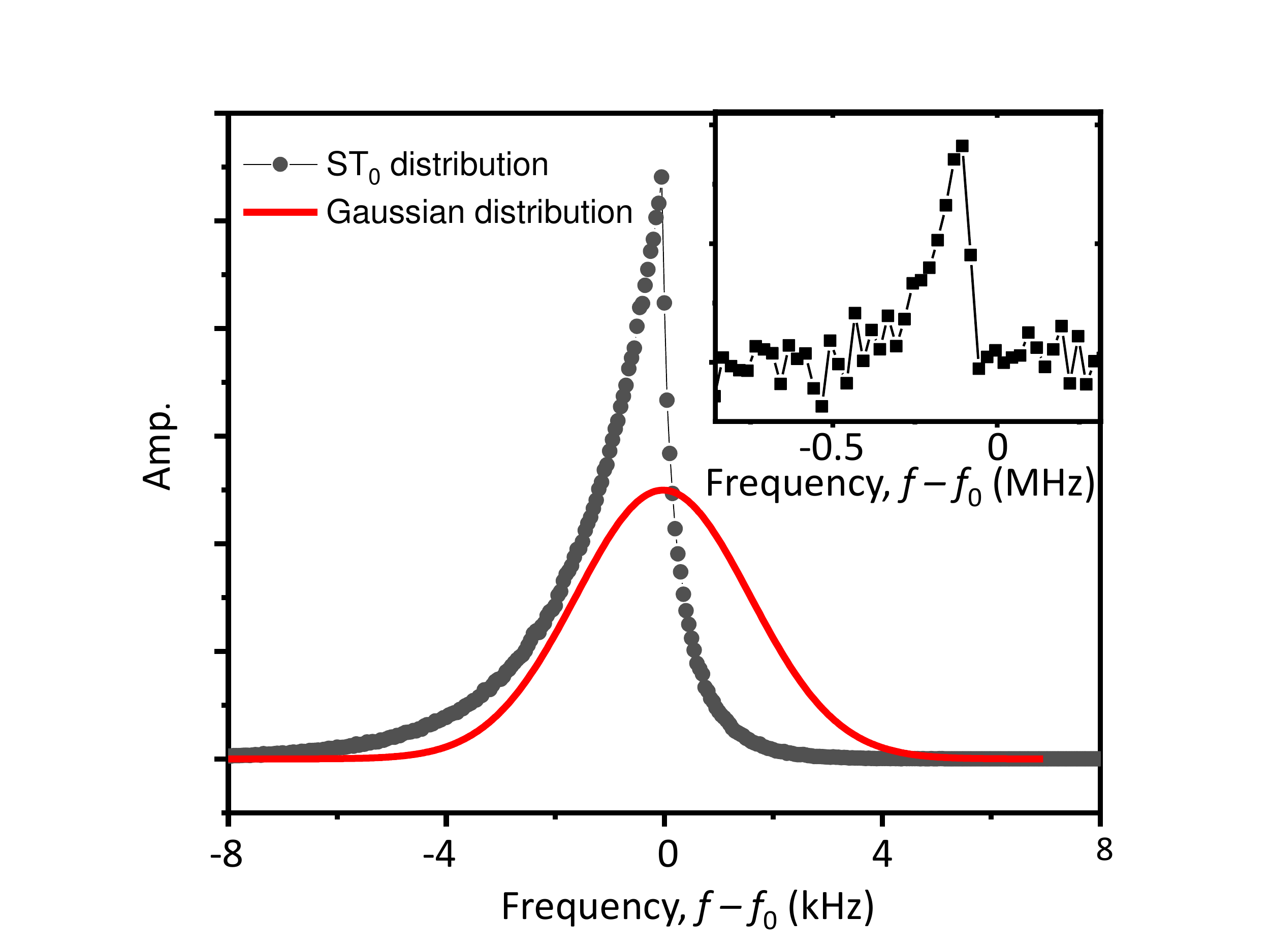} \caption{\textbf{Simulation of the $\text{ST}_0$ spectrum.}
Black points are numerically calculated distribution of the $\text{ST}_{0}$ transition frequencies, while red line is a Gaussian distribution with the same standard deviation. Inset is the experimental spectrum with extra magnetic noise, generated by applying noise currents to the Helmholtz coils. This magnetic noise obeys the normal distribution with standard deviation of 2.3 MHz.
}
\label{Linewidth}
\end{figure}